\documentclass[useAMS,usenatbib]{mn2e}

\usepackage{graphicx}
\usepackage{amsmath}
\usepackage{amssymb}
\usepackage{dcolumn}

\usepackage{longtable}

\usepackage{color}
\usepackage{soul}

\definecolor{red}{rgb}{1.0, 0.0, 0.0}
\definecolor{modifs}{rgb}{1.0, 0.0, 1.0}
\definecolor{darkblue}{rgb}{0.0, 0.0, 0.60}

\newcommand{\red}[1]{\textcolor{red}{\bf #1}}

\usepackage{hyperref}
\hypersetup{bookmarks=true, colorlinks=true, linkcolor=darkblue, citecolor=darkblue, filecolor=darkblue, urlcolor=darkblue}

\defcitealias{2016A&A...592A..84F}{F2016} 
\defcitealias{2017MNRAS.465.2432G}{Paper I}
\defcitealias{2015A&A...575A..34M}{M2015}

\newcommand{\mcc}[1]{\multicolumn{1}{c}{#1}}

\title[MiMeS: the O-type star population]{The MiMeS survey of Magnetism in Massive Stars:\\ 
Magnetic properties of the O-type star population
}

\author[Petit et al.]{V. Petit$^{1}$\thanks{E-mail: VPetit@UDel.edu},
G. A. Wade$^{2}$, F. R. N. Schneider$^{3,4}$, L. Fossati$^{5}$, K. Kamp$^{6}$, \and C. Neiner$^{7}$, A. David-Uraz$^1$, E. Alecian$^{8}$
\\
$^{1}$ Dept. of Physics \& Astronomy, University of Delaware, Newark, DE 19716\\
$^{2}$ Dept. of Physics \& Space Science, Royal Military College of Canada, PO Box 17000, Kingston, ON K7K 7B4, Canada\\
$^{3}$ Zentrum f\"{u}r Astronomie der Universit\"{a}t Heidelberg, Astronomisches Rechen-Institut, M\"{o}nchhofstr. 12-14, 69120 Heidelberg, Germany\\
$^{4}$ Heidelberger Institut f\"{u}r Theoretische Studien, Schloss-Wolfsbrunnenweg 35, 69118 Heidelberg, Germany\\
$^{5}$ Space Research Institute, Austrian Academy of Sciences, Schmiedlstrasse 6, 8042 Graz, Austria\\
$^{6}$ Dept. of Aerospace, Physics and Space Sciences, Florida Institute of Technology, 150 W. University Blvd, Melbourne, FL 32901, USA \\
$^{7}$ LESIA, Paris Observatory, PSL University, CNRS, Sorbonne Université, Université de Paris, 5 place Jules Janssen, 92195 Meudon, France\\
$^{8}$ Univ. Grenoble Alpes, CNRS, IPAG, 38000 Grenoble, France\\
}

\begin{document}
%
%  These Macros are taken from the AAS TeX macro package version 4.0.
%  Include this file in your LaTeX source only if you are not using
%  the AAS TeX macro package and need to resolve the macro definitions
%  in the BibTeX entries returned by the ADS abstract service.
%
%  For more information on the AASTeX macro package, please see the URL
%	http://www.aas.org/publications/aastex.html
%  For more information about ADS abstract server, please see the URL
%	http://adswww.harvard.edu/ads_abstracts.html
%

% Abbreviations for journals.  The object here is to provide authors
% with convenient shorthands for the most "popular" (often-cited)
% journals; the author can use these markup tags without being concerned
% about the exact form of the journal abbreviation, or its formatting.
% It is up to the keeper of the macros to make sure the macros expand
% to the proper text.  If macro package writers agree to all use the
% same TeX command name, authors only have to remember one thing, and
% the style file will take care of editorial preferences.  This also
% applies when a single journal decides to revamp its abbreviating
% scheme, as happened with the ApJ (Abt 1991).

\def\jnl@style{\it}
%commente par Seb
\def\aaref@jnl#1{{\jnl@style#1}}
%ref remplace par aaref pour eviter conflit...

\def\aaref@jnl#1{{\jnl@style#1}}

\def\aj{\aaref@jnl{AJ}}                   % Astronomical Journal
\def\araa{\aaref@jnl{ARA\&A}}             % Annual Review of Astron and Astrophys
\def\apj{\aaref@jnl{ApJ}}                 % Astrophysical Journal
\def\apjl{\aaref@jnl{ApJ}}                % Astrophysical Journal, Letters
\def\apjs{\aaref@jnl{ApJS}}               % Astrophysical Journal, Supplement
\def\ao{\aaref@jnl{Appl.~Opt.}}           % Applied Optics
\def\apss{\aaref@jnl{Ap\&SS}}             % Astrophysics and Space Science
\def\aap{\aaref@jnl{A\&A}}                % Astronomy and Astrophysics
\def\aapr{\aaref@jnl{A\&A~Rev.}}          % Astronomy and Astrophysics Reviews
\def\aaps{\aaref@jnl{A\&AS}}              % Astronomy and Astrophysics, Supplement
\def\azh{\aaref@jnl{AZh}}                 % Astronomicheskii Zhurnal
\def\baas{\aaref@jnl{BAAS}}               % Bulletin of the AAS
\def\jrasc{\aaref@jnl{JRASC}}             % Journal of the RAS of Canada
\def\memras{\aaref@jnl{MmRAS}}            % Memoirs of the RAS
\def\mnras{\aaref@jnl{MNRAS}}             % Monthly Notices of the RAS
\def\pra{\aaref@jnl{Phys.~Rev.~A}}        % Physical Review A: General Physics
\def\prb{\aaref@jnl{Phys.~Rev.~B}}        % Physical Review B: Solid State
\def\prc{\aaref@jnl{Phys.~Rev.~C}}        % Physical Review C
\def\prd{\aaref@jnl{Phys.~Rev.~D}}        % Physical Review D
\def\pre{\aaref@jnl{Phys.~Rev.~E}}        % Physical Review E
\def\prl{\aaref@jnl{Phys.~Rev.~Lett.}}    % Physical Review Letters
\def\pasp{\aaref@jnl{PASP}}               % Publications of the ASP
\def\pasj{\aaref@jnl{PASJ}}               % Publications of the ASJ
\def\qjras{\aaref@jnl{QJRAS}}             % Quarterly Journal of the RAS
\def\skytel{\aaref@jnl{S\&T}}             % Sky and Telescope
\def\solphys{\aaref@jnl{Sol.~Phys.}}      % Solar Physics
\def\sovast{\aaref@jnl{Soviet~Ast.}}      % Soviet Astronomy
\def\ssr{\aaref@jnl{Space~Sci.~Rev.}}     % Space Science Reviews
\def\zap{\aaref@jnl{ZAp}}                 % Zeitschrift fuer Astrophysik
\def\nat{\aaref@jnl{Nature}}              % Nature
\def\iaucirc{\aaref@jnl{IAU~Circ.}}       % IAU Cirulars
\def\aplett{\aaref@jnl{Astrophys.~Lett.}} % Astrophysics Letters
\def\apspr{\aaref@jnl{Astrophys.~Space~Phys.~Res.}}
                % Astrophysics Space Physics Research
\def\bain{\aaref@jnl{Bull.~Astron.~Inst.~Netherlands}} 
                % Bulletin Astronomical Institute of the Netherlands
\def\fcp{\aaref@jnl{Fund.~Cosmic~Phys.}}  % Fundamental Cosmic Physics
\def\gca{\aaref@jnl{Geochim.~Cosmochim.~Acta}}   % Geochimica Cosmochimica Acta
\def\grl{\aaref@jnl{Geophys.~Res.~Lett.}} % Geophysics Research Letters
\def\jcp{\aaref@jnl{J.~Chem.~Phys.}}      % Journal of Chemical Physics
\def\jgr{\aaref@jnl{J.~Geophys.~Res.}}    % Journal of Geophysics Research
\def\jqsrt{\aaref@jnl{J.~Quant.~Spec.~Radiat.~Transf.}}
                % Journal of Quantitiative Spectroscopy and Radiative Transfer
\def\memsai{\aaref@jnl{Mem.~Soc.~Astron.~Italiana}}
                % Mem. Societa Astronomica Italiana
\def\nphysa{\aaref@jnl{Nucl.~Phys.~A}}   % Nuclear Physics A
\def\physrep{\aaref@jnl{Phys.~Rep.}}   % Physics Reports
\def\physscr{\aaref@jnl{Phys.~Scr}}   % Physica Scripta
\def\planss{\aaref@jnl{Planet.~Space~Sci.}}   % Planetary Space Science
\def\procspie{\aaref@jnl{Proc.~SPIE}}   % Proceedings of the SPIE

\let\astap=\aap
\let\apjlett=\apjl
\let\apjsupp=\apjs
\let\applopt=\ao

%-----
\date{
Accepted 2019 September 2. Received 2019 August 23; in original form 2019 June 15}
\pagerange{\pageref{firstpage}--\pageref{lastpage}} \pubyear{2019}
\maketitle
\label{firstpage}
%-----

%-----------------------
\begin{abstract}

In this paper, we describe an analysis of the MiMeS Survey of O-type stars to explore the range of dipolar field strengths permitted by the polarisation spectra that do not yield a magnetic detection. We directly model the Stokes V profiles with a dipolar topology model using Bayesian inference. 
The noise statistics of the Stokes V profiles are in excellent agreement with those of the null profiles. 
Using a Monte-Carlo approach we conclude that a model in which all the stars in our sample were to host a 100\,G, dipolar magnetic field can be ruled out by the MiMeS data.  
Furthermore, if all the stars with no detection were to host a magnetic field just below their detection limit, the inferred distribution in strength of these undetected fields would be distinct from the known distribution in strength of the known magnetic O-type stars. This indicates that the Initial magnetic (B-)field Function (IBF) is likely bimodal -- young O-type stars are expected to either have weak/absent magnetic fields, or strong magnetic fields.
We also find that better upper limits, by at least a factor of 10, would have been necessary to rule out a detection bias as an explanation for the apparent lack of evolved main-sequence magnetic O-type stars reported in the literature, and we conclude that the MiMeS survey cannot confirm or refute a magnetic flux decay in O-type stars. 

\end{abstract}

%-----------------------

%%%%%%%%%%%%
%%%%%%%%%%%%
\section{Motivation and goals}\label{sec:intro}

 Magnetic fields are now routinely detected at the surfaces of a sub-sample of OB stars. These fields generally have a simple, mostly dipolar topology. They are also strong, with dipolar strength values ranging from a few hundred G to tens of kG \citep[e.g.][]{2013MNRAS.429..398P, 2018MNRAS.475.5144S}. Their rarity, along with the lack of observed short-term (days -- decades) evolution, suggests that these fields are not generated by a contemporaneous dynamo mechanism, as is the case for solar-type stars, but are rather a relic from an event or an evolutionary phase that occurred earlier in their past \citep[e.g.][]{1982ARA&A..20..191B}.

 These so-called fossil fields are now increasingly well characterised thanks to the advent of large surveys performed with high-resolution spectropolarimetric instrumentation \citep[e.g.][]{2016MNRAS.456....2W,2015A&A...582A..45F}. These instruments  allow for the measurement of the polarisation change across spectral lines introduced by the Zeeman effect \citep[for an in-depth review, see][]{2009ARA&A..47..333D}. 

 This paper constitutes the second part of the study initiated by \citet[][hereafter Paper I]{2017MNRAS.465.2432G} that characterised the 97 O-type star systems observed with spectropolarimetry within the context of the Magnetism in Massive Stars (MiMeS) survey. 
The MiMeS survey was performed through Large Program allocations with the high-resolution spectrographs ESPaDOnS at the Canada-France Hawaii Telescope, Narval at the T\'elescope Bernard Lyot in France, and HARPSpol at the ESO La Silla 3.6m in Chile \citep{2016MNRAS.456....2W}. 

 \citetalias{2017MNRAS.465.2432G} presented the detection statistics for the survey based on the presence of a polarisation signal in Stokes V, and on the associated measurement of the surface averaged longitudinal field $B_l$ that represents the component of the magnetic field along the line-of-sight, integrated over the stellar disk (thus weighted by the local intensity). 
 
 Among the main results, \citetalias{2017MNRAS.465.2432G} reported that:
 
\noindent (i) the median MiMeS survey precision for the longitudinal field is 50\,G; 

\noindent (ii) the detection of 6 new magnetic O-type stars implies a magnetic incidence fraction of $7\pm3$ percent. In addition, marginal signal was detected in 3 stars that were therefore deemed to be magnetic candidates;

\noindent (iii) there are no correlations between the presence of a magnetic field and any stellar parameters of the sample (other than the expected slow bulk rotation);

\noindent (iv) there is a direct link between the so-called Of?p spectral peculiarities \citep{1972AJ.....77..312W} and magnetism, although not all magnetic O-type stars have these peculiarities.

 In this paper, we perform an analysis of the O-type stars without a magnetic field detection to explore the dipolar field strengths permitted by the polarisation spectra. We directly model the Stokes V profiles with the Bayesian inference procedure developed by \citet{2012MNRAS.420..773P} for a dipolar field topology.

Such a study allows us to address several questions concerning the distribution of magnetic fields in O-type stars, described in the following two sections. It also provides an independent evaluation of the polarisation signature of the magnetic candidate stars, as well as the potential to identify additional magnetic candidates.

%%%%%%%%%%%%
\subsection{Is there a magnetic desert for O-type stars?}
\label{sec:intro-desert}

 In magnetic A-type stars, there is a known deficit of stars with dipole field strengths at the $\sim100$G level \citep{2007A&A...475.1053A}. This so-called ``magnetic desert'' \citep{2014IAUS..302..338L} is a well established result for a few reasons:

\noindent(i) It has been long established that strong magnetism and the Ap-type chemical peculiarities are linked \citep{2007A&A...475.1053A}, making strongly magnetic A-type stars easily identifiable through spectroscopic peculiarities and photometric/spectroscopic variations due to abundance spots. 

\noindent(ii) The field detection limits for A-type stars are generally very good, due to the multitude and sharpness of their metallic spectral features. This has been recently exemplified by the detection of very weak (at the sub-gauss level) magnetic fields at the surface of some A-type stars \citep{2010A&A...523A..41P,2011A&A...532L..13P, 2016A&A...586A..97B}. Clearly, significantly stronger fields would have been noticed -- nevertheless, there is only a handful of known main sequence magnetic A-type stars in the 1-100\,G field strength bin \citep[][]{2016A&A...589A..47A, Blazere2018}. 

 Is there an extension of the ``magnetic desert'' phenomenon for the O-type stars? Using our sample we determine whether the detection limits obtained for O-type stars are sufficient to establish the existence of a magnetic desert in the O-type stars. 

%%%%%%%%%%%%
\subsection{Can we detect magnetic fields in old main sequence O-type stars, and thereby constrain models of the evolution of their surface magnetism?}

 An interesting additional application for this sample of upper limits on dipolar field strength is to explore the cause of the apparent scarcity of evolved (hereafter referred to as ``old'') massive stars (near or past the TAMS) with a detected magnetic field \citep[][hereafter F2016]{2016A&A...592A..84F}.
 \citetalias{2016A&A...592A..84F} compared the age distribution of the known magnetic OB stars with that of a magnitude-limited comparison sample of OB stars. 
These ages were derived using the Bayesian tool \textsc{bonnsai} \citep{2014A&A...570A..66S,2017A&A...598A..60S}, using the evolution tracks of \citet{2011A&A...530A.115B}.
 \citetalias{2016A&A...592A..84F} found that in general, the main sequence fractional ages of the magnetic OB stars are smaller than those of their non-magnetic counterparts, and possibly even more so for stars with higher ZAMS masses.  

 \citetalias{2016A&A...592A..84F} discuss three possible physical explanations for this result:

\noindent (i) The magnetic fields at the surfaces of these stars may decrease more quickly than expected from the magnetic flux conservation hypothesis \citep[the reduction of surface field coming from flux freezing combined with the evolutionary increase in radius;][]{2008A&A...481..465L}, implying a loss of magnetic flux by, for example, Ohmic decay.

\noindent (ii) Magnetic stars could appear younger than they really are when their ages are estimated using non-magnetic evolutionary models, due e.g. to the suppression of convective core-overshoot \citep{2012MNRAS.427..483B}. \citetalias{2016A&A...592A..84F} suggest that in this example case, one would expect an inverse correlation between inferred young apparent age and field strength, which was not observed in their study. 
We note here that other processes involving the interaction between surface magnetic fields and mass-loss (such as magnetic braking and mass-loss quenching) have also been shown to have marked but complex evolutionary effects that should be taken into account when estimating evolutionary ages, especially for more massive stars \citep{2005A&A...440.1041M, 2017MNRAS.466.1052P, 2017A&A...599L...5G,2019MNRAS.485.5843K}.

\noindent (iii) Finally, one proposed pathway for the generation of fossil magnetic fields is via binary mergers \citep[e.g.][]{2016MNRAS.457.2355S}. The rejuvenation of the merger product could lead to a seemingly younger magnetic population. However \citetalias{2016A&A...592A..84F} argue that in a scenario of constant star formation history, the population of mass gainers will be biased toward larger fractional MS age. 

 Importantly, \citetalias{2016A&A...592A..84F} also point out some possible observational biases that could result in the inferred age distribution for known magnetic stars. They performed a preliminary assessment of the magnetic detection limits, based on the field strengths of the weakest known magnetic OB stars as a function of V magnitude. As their estimate suggested that some of the known magnetic stars would still be detectable with current instrumentation upon reaching their TAMS. {As a consequence, observational bias was not considered by those authors to be a likely explanation for the apparent lack of old magnetic OB stars.} However, this assessment of observational bias does not take into account the characteristics of the typical spectropolarimetric survey -- some of the known magnetic stars were first identified by other means (e.g. indirect magnetospheric emission/variation), and may have been observed with deeper detection limits than that of the typical large survey. 

 With the MiMeS O-star sample, we can directly address the observational biases, for O-type stars, by answering the two following questions:
\begin{itemize}
\item Has a sufficiently large sample of  old MS O-stars been searched for surface magnetic fields?
\item If so, were the detection limits sufficiently precise to inform models of the evolution of their surface magnetism?
\end{itemize}

%%%%%%%%%%%%
\subsection{Paper structure}

 The paper is structured in the following way.
 In \S\ref{sec:obs}, we briefly summarise the observations. 
 In \S\ref{sec:bayes}, we summarise the application of the Bayesian technique of \citet{2012MNRAS.420..773P} to our sample, including the way in which we treat spectroscopic binary systems. 
 In \S\ref{sec:results}, we present the statistical properties of our sample of magnetic field upper limits. 
 In \S\ref{sec:discussion}, we interpret our results in the context of (i) evaluating the presence of a magnetic desert in magnetic O-type stars, and (ii) explaining the apparent lack of old, main sequence magnetic massive stars. 
 We finally summarise our findings in \S\ref{sec:conclusion}.

%%%%%%%%%%%%
%%%%%%%%%%%%
\section{Observations} \label{sec:obs}

 The spectropolarimetric observations used here and in \citetalias{2017MNRAS.465.2432G} were obtained in the larger context of the Magnetism in Massive Stars (MiMeS) survey, described by \citet{2016MNRAS.456....2W}. To summarise, high-resolution echelle spectropolarimetric observations enable the measurement of the change in polarisation across spectral lines induced by the Zeeman effect. 

 The sample of \citetalias{2017MNRAS.465.2432G} consists of 432 nightly-averaged Stokes V observations of 97 O-type stars. 
 Our sample here consists of stars without definite magnetic detection, reducing it to the 91 stars listed in Table \ref{tab:stars} (which additionally contains the 11 O-type secondary stars in SB2/SB3 system discussed in \S \ref{sec:bin}). Column 1 provides an identification number for easy reference, column 2 the HD number or BD number, and column 3 the spectral type. 

In \citetalias{2017MNRAS.465.2432G}, the Least-Squares Deconvolution (LSD) method \citep{1997MNRAS.291..658D} was used to produce a high signal-to-noise, mean line profile from each nightly-averaged observation. 
The underlying principle of this method is that in the limit where the Zeeman splitting has a minimal impact on the intensity line profile \citep[the so-called weak-field approximation;][]{2004ASSL..307.....L}, the shape of the Stokes V profile remains the same from line to line, with an amplitude scaled by the wavelength and effective Land\'e factor of the electronic transition. 
We use these same LSD profiles here. Within the detection limits obtained in our results, the weak-field approximation is always valid. Column 4 of Table \ref{tab:stars} gives the number of nightly-averaged observations per star. 

 In \citetalias{2017MNRAS.465.2432G}, each nightly-averaged LSD intensity profile was modelled to measure $v\sin i$. For the 36 stars for which more than one observation is available, the standard deviation of the $v\sin i$ values reported in \citetalias{2017MNRAS.465.2432G} are less than 10 percent of the mean $v\sin i$ value, with the exception of HD\,167262. A visual inspection of the LSD profiles for this star does not show significant variation in total line broadening -- the lower values of $v\sin i$ reported in \citetalias{2017MNRAS.465.2432G} were accompanied with larger values of macroturbulence velocities. 
 Column 5 of Table \ref{tab:stars} lists, for each star, the $v\sin i$ averaged over all observations that we use in our modelling in \S\ref{sec:bayes}.

%%%%%%%%%%%%
%%%%%%%%%%%%
\section{Bayesian modelling}\label{sec:bayes}

 We compared the LSD profiles of each star to a grid of synthetic Stokes $V$ profiles using the method of \citet{2012MNRAS.420..773P}. 
In this approach, we assume a simple centred dipolar field model $M_{B_\mathrm{pole}}$, parametrised by the dipole field strength $B_\mathrm{pole}$, the rotation axis inclination $i$ with respect to the line of sight, the positive magnetic axis obliquity $\beta$, and a set of rotational phases $\Phi=[\varphi_1..\varphi_N]$ associated with a set of Stokes $V$ observations of a single star. 

 Given that the rotation periods of the survey stars are effectively unknown, we treat the rotational phases as nuisance parameters to obtain the goodness-of-fit of a given rotation-independent $\mathcal{B}$=[$B_\mathrm{pole}$, $i$, $\beta$] magnetic configuration in a Bayesian statistical framework:
\begin{equation} \label{eq|post_conf}
	p(\mathcal{B} | \mathcal{D}, M_{B_\mathrm{pole}}) = \int \frac{p(\mathcal{B}, \Phi | M_{B_\mathrm{pole}})p(\mathcal{D} | \mathcal{B}, \Phi, M_{B_\mathrm{pole}})}{p(\mathcal{D} | M_{B_\mathrm{pole}})} \mathrm{d}\Phi. 
\end{equation}

 The Bayesian prior for the inclination is described by a random orientation $[p(i|M_1) = \sin(i)\,di]$. 
Given that the inclination angle cannot be strictly 0$^\circ$ if rotational broadening is present ($v\sin i>0$), we restrict the range of inclination angle values explored to angles larger than $i > \sin^{-1}(v_\mathrm{eq}\sin i / v_\mathrm{eq, max})$. We can estimate $v_\mathrm{eq, max}$ by the break-up velocity of the stars ($V_\mathrm{crit}=\sqrt{GM/1.5R_\mathrm{pole}}$). We adopt a conservatively overestimated value of 700\,km\,s$^{-1}$ for the whole sample. A overestimation of $v_\mathrm{eq, max}$ leads to the inclusion of smaller inclinations in the parameter space. This in turn gives less weight to the inclination angles around 90 degree, which are more likely to produce a low amplitude or no Stokes V signal. We may therefore slightly overestimate the magnetic field upper limits, although not by much, as the prior probabilities for low-inclination configurations are low. 

The dipole field strength Bp is a parameter that can vary over several decades (from a few dozen G to a few kG). We therefore used a Jeffreys prior, which sets an equal probability per decade and therefore represents a lack of information about the scale of the parameter. To avoid a singularity at $B_\mathrm{pole} = 0$\,G, we used the modified form of \citet{2005ApJ...631.1198G}. The obliquity and the rotational phases have constant priors, appropriate for positional parameters. 

 The synthetic flux profiles are obtained by numerically integrating the emergent intensities over the projected stellar disk of a given magnetic geometric configuration.  We use a limb darkening law of the form $I(\mu)/I(\mu=1)=1-\epsilon+\mu\epsilon$, where $\mu$ is the cosine of the angle between the ray direction and the normal to the stellar surface. As we are using line profiles resulting from the LSD method (with spectral lines covering a broad range of visible wavelengths), we use the fiducial, grey value of $\epsilon=0.6$ for the whole sample. 

 The emergent intensities are calculated using the weak-field approximation described by \citet{2004ASSL..307.....L}. 
For this study we use Voigt-shaped line profiles with a damping constant $a=10^{-2}$ and a thermal speed $v_\mathrm{th}=7$\,km\,s$^{-1}$, appropriate for the stellar temperatures under consideration as the LSD profiles are mostly composed of metallic spectral lines. 

 We adjust the line depth and the isotropic Gaussian macroturbulence\footnote{Of the form $\mathrm{e}^{-v^2/v^2_\mathrm{mac}}/(\sqrt{\upi}v_\mathrm{mac})$.} velocity $v_\mathrm{mac}$ to fit the mean of the Stokes $I$ LSD profiles of a given star. 

 We performed this analysis on the circular polarisation LSD Stokes V profiles as well as the null polarisation LSD profiles.

\subsection{Spectroscopic binaries}
\label{sec:bin}

 We consider that the presence of a non-magnetic spectroscopic binary companion may have two effects (other that the radial velocity shift) on the LSD profiles and their analysis. First, a companion contributes continuum flux to the observed spectrum, reducing the amplitude of any Stokes V profile contributed by the primary star in the normalised circular polarisation spectrum. Simultaneously, the depths of spectral lines of the star in the Stokes I spectrum are also reduced by the same factor. As a consequence, the ratio of 
V to I is preserved, and the net result is only to increase the inferred uncertainty of the magnetic diagnosis. In other words, the continuum flux of a companion increases the upper limit on an undetected field, but does not introduce any systematic error.

The second effect results from the presence of the spectral lines of the companion. In this case, the companion's lines may blend with those of the studied star, making the determination of the relevant velocity range for the diagnosis of Stokes V profiles ambiguous. In addition, spectral line blending contributes to the equivalent width of Stokes I, but has no influence on the Stokes V profile, assuming the companion is not magnetic. Hence uncorrected line blending of a companion may systematically affect the magnetic diagnosis. In cases where we might have failed to recognise a SB2 system and treated it as a single star, we would underestimate the maximum field strength allowed by the data for the primary by at most a factor of 2, which corresponds to the extreme case of two O-type stars with identical spectral type and luminosity.

Our sample contains 20 SB2 systems, for which we correct for the line blending effect by modelling the components' contributions to the Stokes I profile, as described in \citetalias{2017MNRAS.465.2432G}. These stars are identified in Table \ref{tab:stars} with a `SB2' label in column 1.

 As our modelling requires a single set of line parameters for all observations, we find the best fit to the ensemble of observations that can be obtained by only adjusting the radial velocities. The fits are presented in Appendix \ref{sec:apSB2}.

 We perform our analysis as above by making the assumption that the secondary star does not have a magnetic field. 
This is reasonable as no Stokes V signal has been definitively detected in any stars in the sample adopted for analysis in this study.
Therefore, to be more precise, this analysis does not take into account the possibility that the polarisation from both stars cancels out -- which  would be relevant only for observations at an orbital phase when the spectral lines overlap significantly.

 For 10 stars within the binary sample, the literature indicates that the secondary is also an O-type star. For these stars, we perform our calculation again, considering that the primary is not magnetic and the secondary might be magnetic. This makes the assumption that the mask was suitable for both sets of spectral lines -- another reason to only do these calculations for O-type secondaries. 
In one case, HD 17505, the spectral lines from 3 O-stars are present: both companions were included.

\subsection{Odds ratios}

 To assess the presence of a dipole-like signal in our observations, we compute the odds ratio of the null model ($M_{B_\mathrm{pole}=0}$; no magnetic field implying Stokes $V = 0$) with the dipole model ($M_{B_\mathrm{pole}}$):
\begin{equation}
	\frac{M_{B_\mathrm{pole}=0}}{M_{B_\mathrm{pole}}} = \frac{p(M_{B_\mathrm{pole}=0})}{p(M_{B_\mathrm{pole}})} \frac{p(\mathcal{D}|B_\mathrm{pole}=0)}{p(\mathcal{D}|M_{B_\mathrm{pole}})}.
\end{equation}
The resultant metric, $\log(M_{B_\mathrm{pole}=0}/M_{B_\mathrm{pole}})$, is displayed in Table~\ref{tab:stars} for Stokes $V$ (column 6) and for the null $N$ profiles (column 7).

 In general, the odds ratios are in favour of the non-magnetic model by approximately half an order of magnitude. Note that both the magnetic and the non-magnetic model can reproduce a signal consisting of only pure noise equally well, as the case $B_\mathrm{pole}=0$\,G is a valid parameter of the magnetic model. 
 Therefore in the case of pure noise the difference between the two models is expected to be dominated by the ratio of priors that penalises the magnetic model for its added complexity. Therefore, in this nested model we consider any odds ratios not strongly in favour of the magnetic model to indicate the absence of significant magnetic signal in the observations. According to \citet{Jeffreys1998}, the evidence in favour of a given model is considered  moderate when $>$$10^{1.0}$ (10:1), strong when $>$$10^{1.5}$ (30:1), and very strong when $>$$10^{2.0}$ (100:1).

\subsection{Probability density function}

As discussed by \citet{2012MNRAS.420..773P}, no meaningful constraint can be placed on the dipole geometry ($i$ and $\beta$) when no circular polarisation signal is present. We therefore treat the dipole geometry parameters as nuisance parameters to extract the probability density distribution for the dipolar field strength:
\begin{equation}
	p(B_\mathrm{pole}|\mathcal{D}, M_{B_\mathrm{pole}}) = \int p(\mathcal{B}|\mathcal{D}, M_{B_\mathrm{pole}})\,\mathrm{d}i\,\mathrm{d}\beta .
\end{equation}

Except for stars with strong evidence in favour of the magnetic model, all the probability distributions generally peak at 0\,G and have an extended tail at large field values, caused by the few dipole orientations that never result in a circular polarisation signal. 
From this probability density, we extract the credible regions around the maximum probability value enclosing a given percentage of the integrated probability. With a Gaussian-shaped probability density, enclosed probabilities of 68.3, 95.4, and 99.7 percent would be analogous to the 1$\sigma$, 2$\sigma$, and 3$\sigma$ contours in frequentist statistics. However, our probability distributions generally deviate strongly from a Gaussian \citep[see discussion by][]{2012MNRAS.420..773P} and these credible regions are generally not linear (i.e. the upper limit of the 99.7 percent credible region for $B_\mathrm{pole}$ will generally reach farther from the peak value than 3 times that of the 68.3 percent credible region).
In Table~\ref{tab:stars} (columns 8-11), we compile the credible region upper limits for the 68.3 and 95.4 percent credible regions, the latter being associated with the field strength that would generally have been detectable in the large majority of field configurations \citep[as demonstrated by][]{2012MNRAS.420..773P}.

\section{Statistical results}\label{sec:results}

\begin{figure*}
	\includegraphics[width=\textwidth]{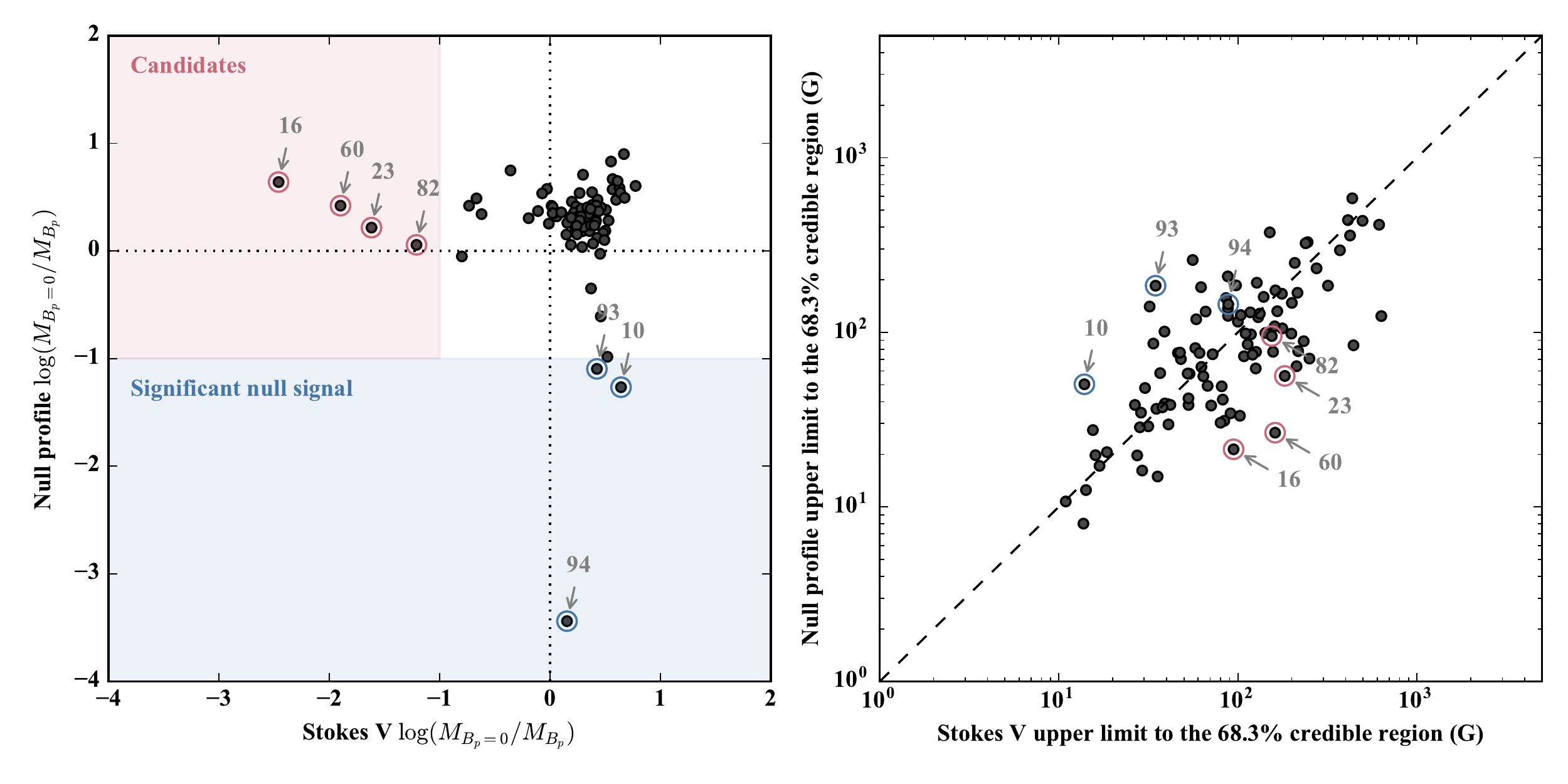}
	\caption{ \label{fig-VNcomparison} \textit{Left}: Comparison of the odds ratios $\log( M_{B_\mathrm{pole}=0} / M_{B_\mathrm{pole}} )$ for the null profiles and the Stokes V profiles. The region below the horizontal line contains stars for which the magnetic model is preferred for the null profiles. The region to the left of the vertical line contains stars for which the magnetic model is preferred for the Stokes V profiles. 
	The blue shaded region indicates odds ratios indicative of significant null signal. The red shaded region indicates the location of magnetic candidates. 
	Specific stars discussed in the text are labeled with their identification number as assigned in Table \ref{tab:stars}. \textit{Right:}  Upper limits of the credible region corresponding to 68.3 percent of the probability, for the null profiles and for the Stokes V profiles. The dashed line shows the one-to-one correspondence. }
\end{figure*}
%----------

 Fig.\,\ref{fig-VNcomparison} (left) shows a comparison of the odds ratios $\log( M_{B_\mathrm{pole}=0} / M_{B_\mathrm{pole}} )$ for the null profiles (on the $y$-axis) and for the Stokes V profiles (on the $x$-axis). 
The region below the horizontal line represents stars for which the magnetic model is preferred for the null profiles. Therefore any stars located in the shaded blue region (more than 10:1 in favour of the magnetic model) are considered to possibly have instrumental spurious signal or stellar variability of non-magnetic origin. 

\begin{figure}
	\includegraphics[width=0.40\textwidth]{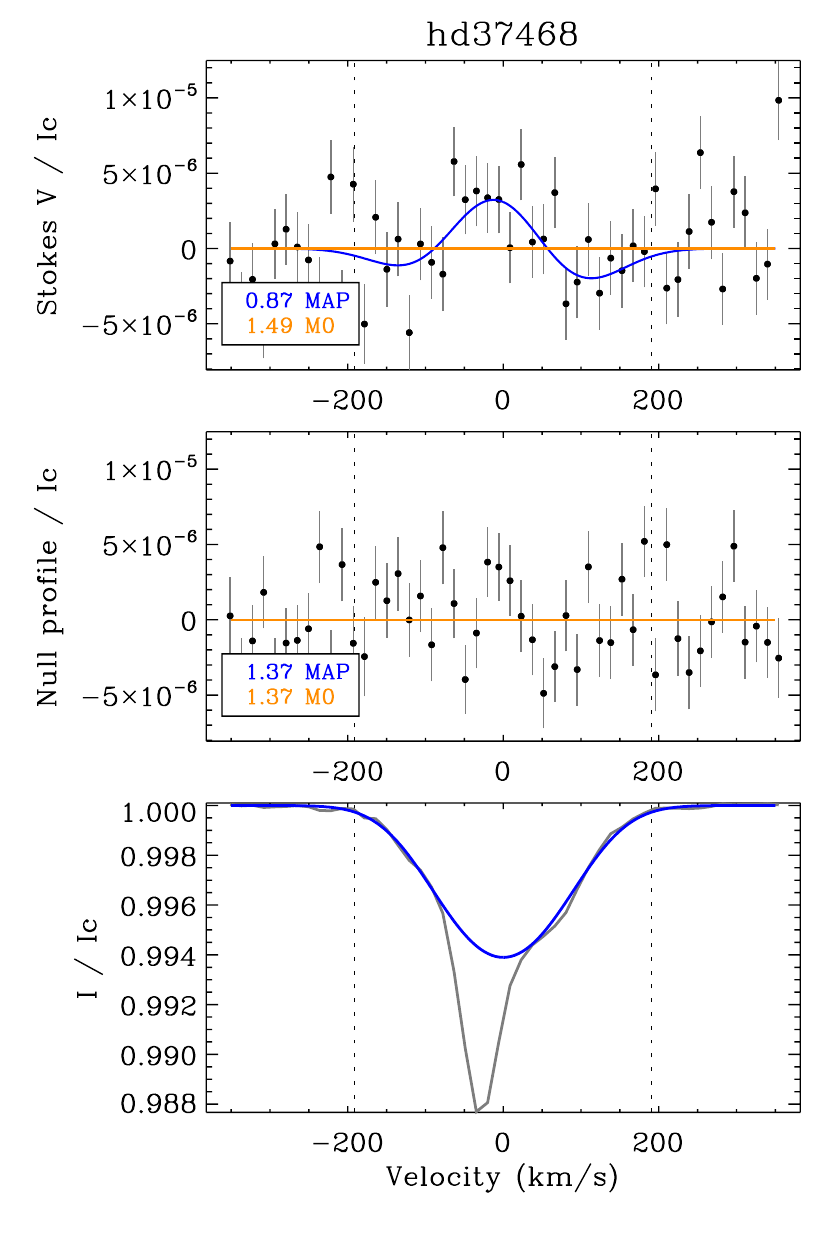}
	\caption{ \label{fig-candidate} LSD profile of the magnetic candidate HD 37468 (\#23). The bottom, middle, and top panels show the intensity, null, and Stokes V profiles, respectively. The non-magnetic model (MO) and the model with the maximum a posteriori probability (MAP) are shown in orange and blue, respectively. The corresponding reduced $\chi^2$ are indicated in the legends. }
\end{figure}

 The region to the left of the vertical line represents stars for which the magnetic model is preferred for the Stokes V profiles. 
Stars with Stokes V profiles yielding an odds ratio at least 10:1 in favour of the magnetic model ($\log( M_{B_\mathrm{pole}=0} / M_{B_\mathrm{pole}} )<-1$) and no evidence of significant null signal (red shaded region) are therefore considered to be magnetic candidates. These candidates are in agreement with those reported in \citetalias{2017MNRAS.465.2432G} (their Table 4 and their Fig.\,4), with the addition of the primary component of HD 37468 (Fig.\,\ref{fig-candidate}). 

In addition to the candidate magnetic stars, \citetalias{2017MNRAS.465.2432G} identifies a list of 6 probable spurious detections in the Stokes V profiles. These stars do not appear as such with our method -- the excess Stokes V signal does not match the expected shape for a dipolar magnetic field yielding an odds ratio inferior to the 10:1 threshold ($\log( M_{B_\mathrm{pole}=0} / M_{B_\mathrm{pole}}>-1 )$). We hence confirm that these are indeed spurious detections.

\citetalias{2017MNRAS.465.2432G} computed two incidence rates of magnetic stars: $5.6\pm2.3$ percent considering only detected magnetic stars and $8.3\pm2.8$ percent considering the magnetic candidates, yielding a mean incidence of $7\pm3$ percent. The addition of one magnetic candidate gives $5.6\pm2.3$ percent and $9.3\pm2.9$ percent, respectively, therefore not modifying the reported $7\pm3$ percent incidence within the error bar. 

\vspace{1cm}

%----------

 The right panel of Fig.\,\ref{fig-VNcomparison} compares the upper limits of the credible region corresponding to 68.3 percent of the probability, again for the null ($y$-axis) and the Stokes V ($x$-axis) profiles. The individual stars for which the odds ratio was strongly in favour of the magnetic model, either for the null profile or the Stokes V profiles, are circled in blue and red, respectively. In general, the upper limits for our sample show a correspondence between the null profiles and Stokes V profiles with a standard deviation of $\sim$0.3 dex. 

 We thus conclude that apart from the 3 magnetic candidate stars identified by \citetalias{2017MNRAS.465.2432G} and the additional magnetic candidate identified in this study, there is no strong evidence for a magnetic signal in any other individual star in our sample. 

We now analyse our sample as a population, to determine which types of magnetic field strength distributions would be compatible with our measurements.

%###################

\subsection{Population analysis}

\label{sec:pop}

%----------
\begin{figure}
	\includegraphics[width=0.5\textwidth]{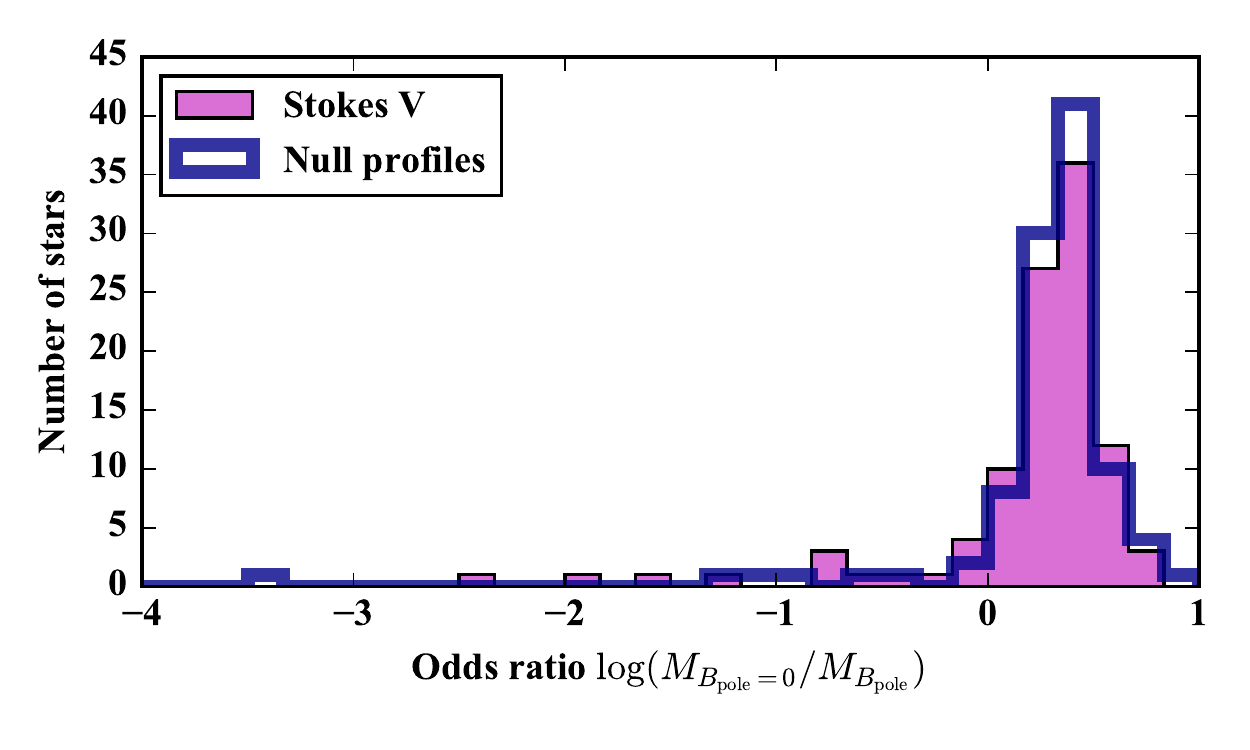}
	\caption{ \label{fig-VN-odds-histo} Distributions of odds ratios derived from the Stokes V profiles and from the null profiles, most likely drawn from the same parent distribution.}
\end{figure}
%----------

In Fig.\,\ref{fig-VN-odds-histo}, we compare the distribution of odds ratios derived from the Stokes V profiles and the null profiles (filled red histogram and empty blue histogram, respectively)\footnote{The inclusion or exclusion of the candidate magnetic stars or of the stars with significant null signal do not change the final conclusions. In the following discussion, we consider the full sample.}. A Kolmogorov-Smirnov (KS) test results in a 69 percent probability that these two distribution are likely drawn from the same parent distribution. This implies that the odds ratios derived from the Stokes V profiles are consistent with the instrumental noise as represented by the null profiles. 

%----------
\begin{figure}
	\includegraphics[width=0.5\textwidth]{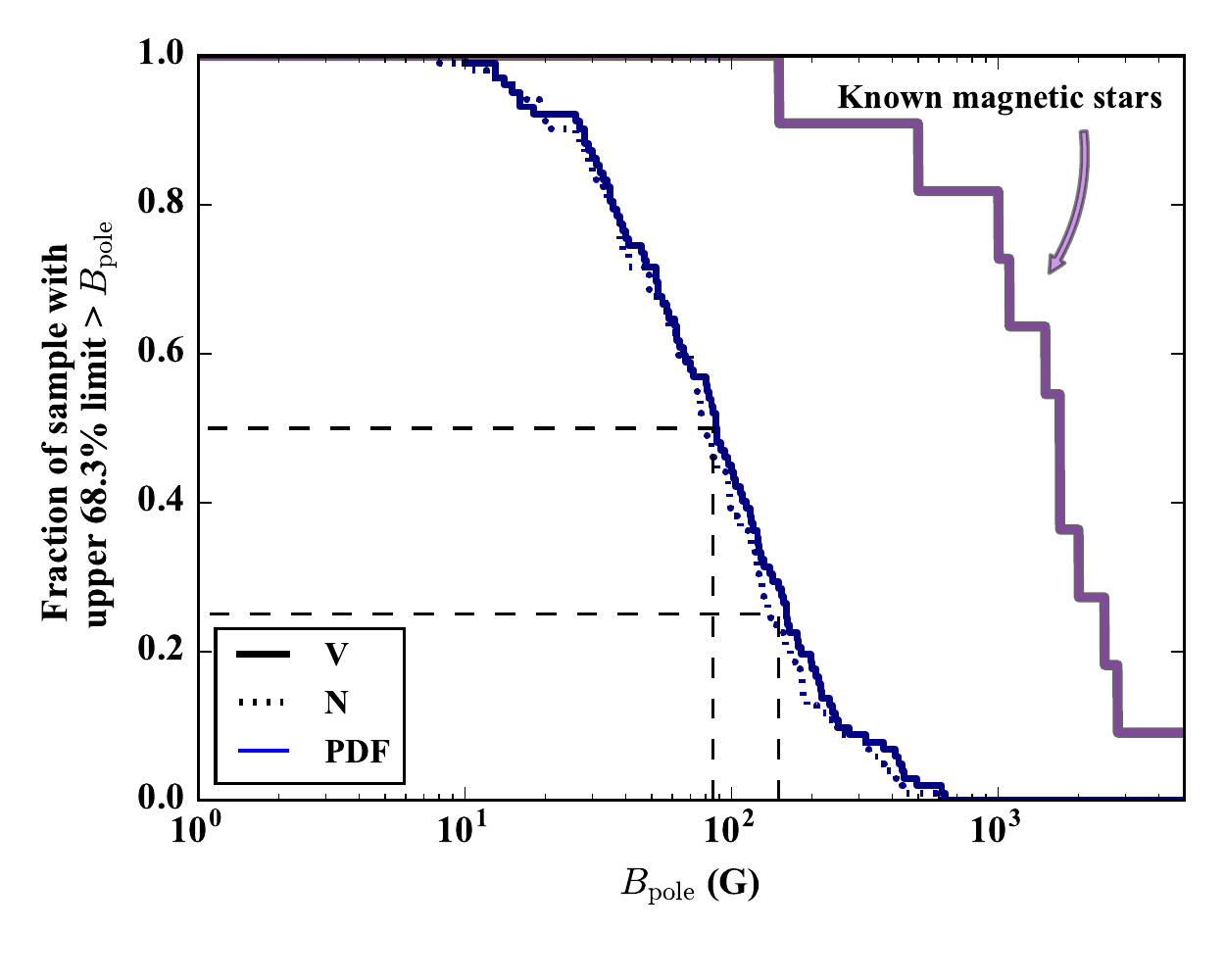}
	\caption{ \label{fig-comp-Mag} Cumulative histograms of the fraction of the sample with an upper limit to the 68.3 percent credible region \textit{larger} than a certain dipole strength value $B_\mathrm{pole}$, for the Stokes V profiles (solid blue) and the null profiles (dotted blue). 
	The black dashed lines illustrate that (i) 25 percent of the stars in our sample have upper limits larger than the lowest, detected magnetic field in an O-type star \citep[$\sim140$\,G;][]{2015A&A...582A.110B}), and (ii) half of the stars in our sample have 68.3 percent upper limits better than 85\,G. 
		The cumulative histogram of the currently known magnetic O-type stars (in pink) is illustrated as the fraction of magnetic stars with a dipolar field strength larger than $B_\mathrm{pole}$.}
	
\end{figure}
%----------

 Fig.\,\ref{fig-comp-Mag} shows cumulative histograms of the fraction of the sample with an upper limit to the 68.3 percent credible region that is larger than a certain dipole strength value. 
The distribution for the Stokes V profiles (solid dark blue) agrees with the null profile distribution (dashed dark blue) -- a KS test results in a $\sim$90 percent probability that these two samples were drawn from the same parent distribution.

We compare these to the distribution of dipolar field strengths of the 11 known magnetic O-type stars, illustrated in Fig.\,\ref{fig-comp-Mag} as the fraction of magnetic stars with a dipolar field strength larger than a certain value\footnote{The magnetic star distribution does not reach zero in our figure, as NGC 1624-2 has a field of 20\,kG, outside the range of our figure.}. We use the magnetic strength values reported in the compilation of \citet{2013MNRAS.429..398P}, with the addition of the  magnetic O-type star HD 54879 subsequently detected by \citet{2015A&A...581A..81C}.
A KS test results in a $10^{-5}$ probability that the distribution of the 68.3 percent upper limits and the distribution of the known magnetic field strengths are drawn from the same distribution. 
Even when considering the upper limits of the 99.7 percent credible region, a KS test also indicates that they are not compatible ($10^{-2}$ probability that they are drawn from the same distribution).

{Thus at first glance, we can conclude that if all the stars in our sample host a magnetic field just below their detection limit, the distribution in strength of these undetected fields would be different from the distribution in strength of the known magnetic O-type stars. This could suggest a bimodal distribution of magnetic fields, grouped in separate -- very weak or very strong -- regimes. However this simple comparison might not be viewed as fair, given the expected evolution of surface fields with age, which we will address in \S\ref{sec:discussion}.}

%----------
\begin{figure*}
	\includegraphics[width=\textwidth]{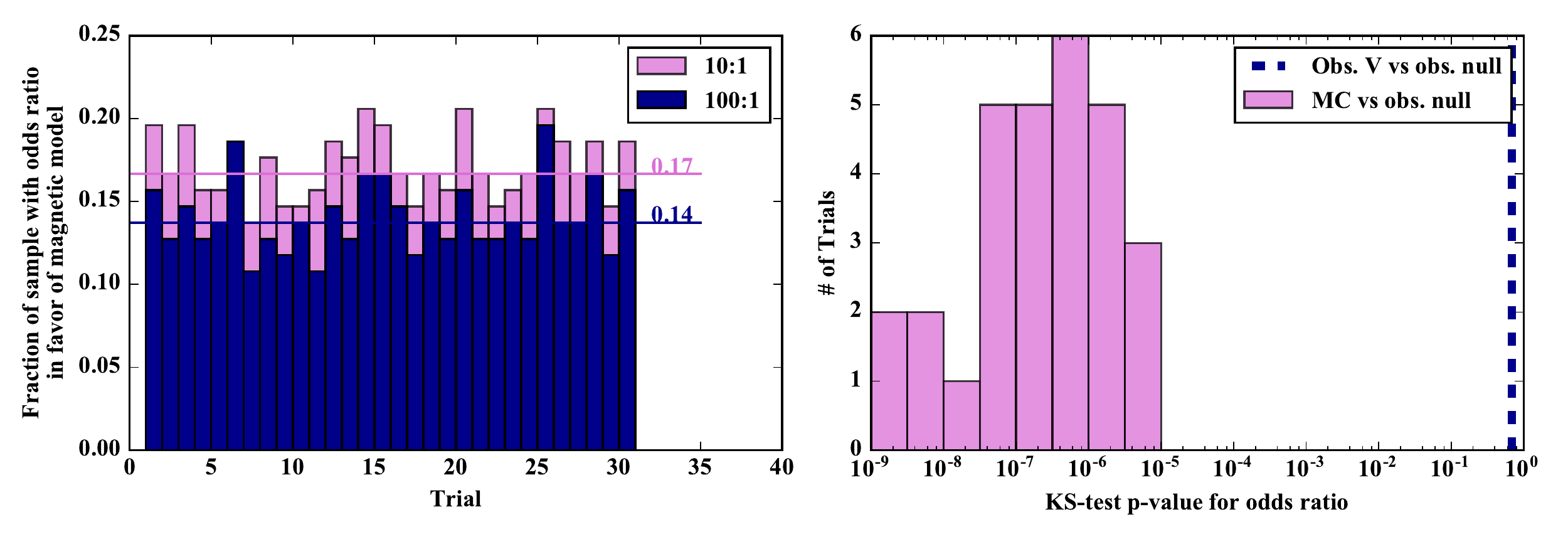}
	\caption{ \label{fig-odd-MC} {Results from 30 Monte Carlo calculations in which the signal from a 100\,G dipolar field with a random orientation was injected into the $N$ profile in every star. \textit{Left}: number of stars for which the odds ratio is greater than 10:1 and 100:1 in favour of the magnetic model, in light pink and dark blue respectively. The mean fraction for each threshold is indicated by the horizontal line of the appropriate colour. \textit{Right:} Histogram of KS-test values calculated between the observed distribution of odds ratios for the original null profile and each Monte-Carlo distribution (in light pink). This is compared with the good agreement between the observed distribution of odds ratio for Stokes $V$ and for the null profile (dark blue dashed line).}}
\end{figure*}
%----------
%----------
\begin{figure*}
	\includegraphics[width=\textwidth]{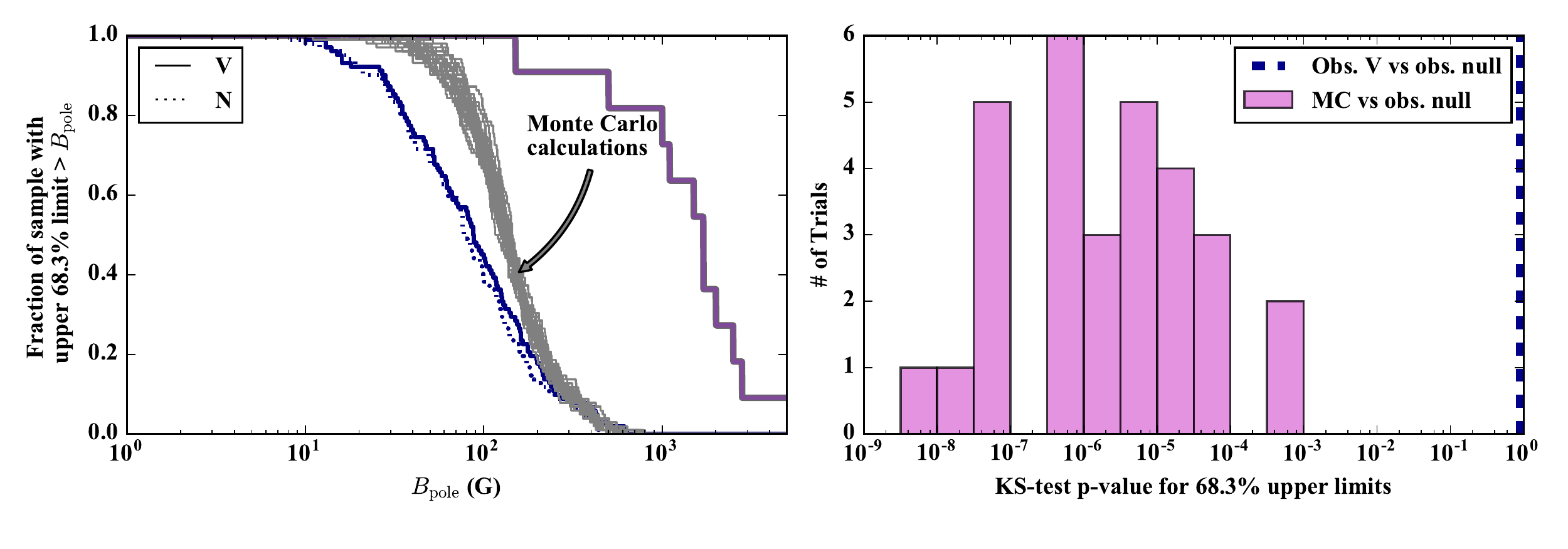}
	\caption{ \label{fig-upper-MC}  {Results from 30 Monte Carlo calculations in which the signal from a 100\,G dipolar field with random orientation was injected into the null profile for every star. \textit{Left}:  Same as Fig.\,\ref{fig-comp-Mag}, with the addition of the distribution of 68.3 percent upper limits for each Monte Carlo calculation (grey histograms). \textit{Right}: Same as the right panel of Fig.\,\ref{fig-odd-MC}, but for the distribution of 68.3 percent upper limits.  }}
\end{figure*}
%----------

 But first, we set out to address the following question: \textbf{could \textit{all} of the O-type stars in the sample have a magnetic field slightly weaker than the current, average observation limits?} As all the stars in our sample have various detection limits due to factors such as brightness, SNR of the observations, line broadening, number of spectral lines available for LSD multi-line technique, etc, the best way to explore this question is through a Monte Carlo simulation.

We perform a Monte-Carlo simulation by calculating sets of synthetic Stokes $V$ signatures corresponding to a dipolar magnetic field with a polar strength of 100~G. This rounded field value corresponds roughly to the upper limit of the 63.8 percent credible region achieved for 50 percent of the sample. It also corresponds to the order of magnitude in field strength at which the magnetic desert is observed. 

For each star, the geometry of the dipolar field ($i$ and $\beta$) and the observational rotational phases ($\Phi$) are randomly selected. All these parameters are drawn from a uniform distribution, apart from the inclination of the rotational axis that is drawn from a $\sin(i)$ distribution (as discussed in \S\ref{sec:bayes}). The characteristics of the intensity profiles are the same as those used for the real sample. The corresponding synthetic magnetic signal is injected directly into the real null profile signal, which is a direct measure of the noise of each individual LSD profile. The Bayesian probability distributions and odds ratios were recalculated using these new profiles. We performed 30 realisations of this Monte-Carlo simulation over our sample.

 Fig.\,\ref{fig-odd-MC} (left) shows the fraction of the sample with an odds ratio more than 10:1 (light pink), and 100:1 (dark blue) in favour of the magnetic model for each Monte Carlo realisation. The unambiguously detected fraction in the simulated sample is $\sim$15 percent. In other words, 15 percent of this synthetic sample (i.e. $\sim15$ stars) of uniformly-magnetized stars would have been detected in the MiMeS survey, while 85 percent would have remained undetected. However, the right panel of Fig.\,\ref{fig-odd-MC} shows a histogram (in light pink) of KS test values between the observed distribution of odds ratios obtained from the null profile and the distribution of odds ratios for each Monte Carlo realisation. In comparison to the good agreement between the observed distribution calculated from the Stokes V and null profiles, as indicated by the high value obtained for the KS test (dark blue dashed line in Fig.\,\ref{fig-odd-MC}), the distribution of odds ratios resulting from the Monte Carlo simulations are significantly different from the observed distribution from the null profiles.  This is also the case when comparing the Monte Carlo distributions to the observed odds ratio distribution obtained from the Stokes V profiles instead of the ones from the null profiles. 
 Fig.\,\ref{fig-upper-MC} (left) also compares the distribution of 68.3 percent credible region upper limits obtained for each Monte Carlo simulation (grey curves) with the observed distributions from the Stokes V and null profiles. Once again, a KS test between these simulated distributions and the observed V and null distributions shows that they are significantly different -- the right panel of Fig.\,\ref{fig-upper-MC} shows the histogram of KS values (in light pink) once again compared to the  KS value between the observed distribution calculated from the V and N profiles (in dark blue). 

\textbf{We therefore conclude that even though the direct detection rate would be rather low -- only slightly higher than the measured bulk incidence of magnetism in massive stars -- the presence of a 100\,G dipolar field in every star in our sample would have resulted in a distribution of odds ratio from the Stokes $V$ profiles that is statistically different from those obtained from the null profiles. In other words, if all the stars had a 100\,G field, we would have noticed.} The same is also true for the distributions of credible region upper limits. 
To test whether this result is simply driven by a small number of very high-quality observations, we also remove from the sample the simulated observations which yield an odds ratio greater than 10:1 in favour of the magnetic model and recalculate the KS tests. The results lead to the same conclusion.

\subsection{Completeness of the MiMeS O-star survey}

Our Monte-Carlo calculations also provide a way to quantify the completeness of the MiMeS survey of O-type stars to the detection of a star with a magnetic field of a certain strength. 
From the case presented in the previous sub-section, injecting a simulated Stokes $V$ signature in every star in our sample only yields an unambiguous detection (from the odds ratios) for 15 percent of the sample. For a 500\,G field, this fraction goes up to 70 percent. 
Let us consider a sample of O-type stars with the same characteristics as the MiMeS sample containing only one magnetic star, with a dipolar strength of 100\,G/500\,G. This 15/70 percent can be thought of as the probability of detecting that magnetic star among the sample. 
In fact, these fractions obtained from the Monte Carlo calculations follow closely the cumulative distribution constructed from the upper limits to the 95.4 percent credible regions for $B_\mathrm{pole}$. 

\begin{figure}
	\includegraphics[width=0.5\textwidth]{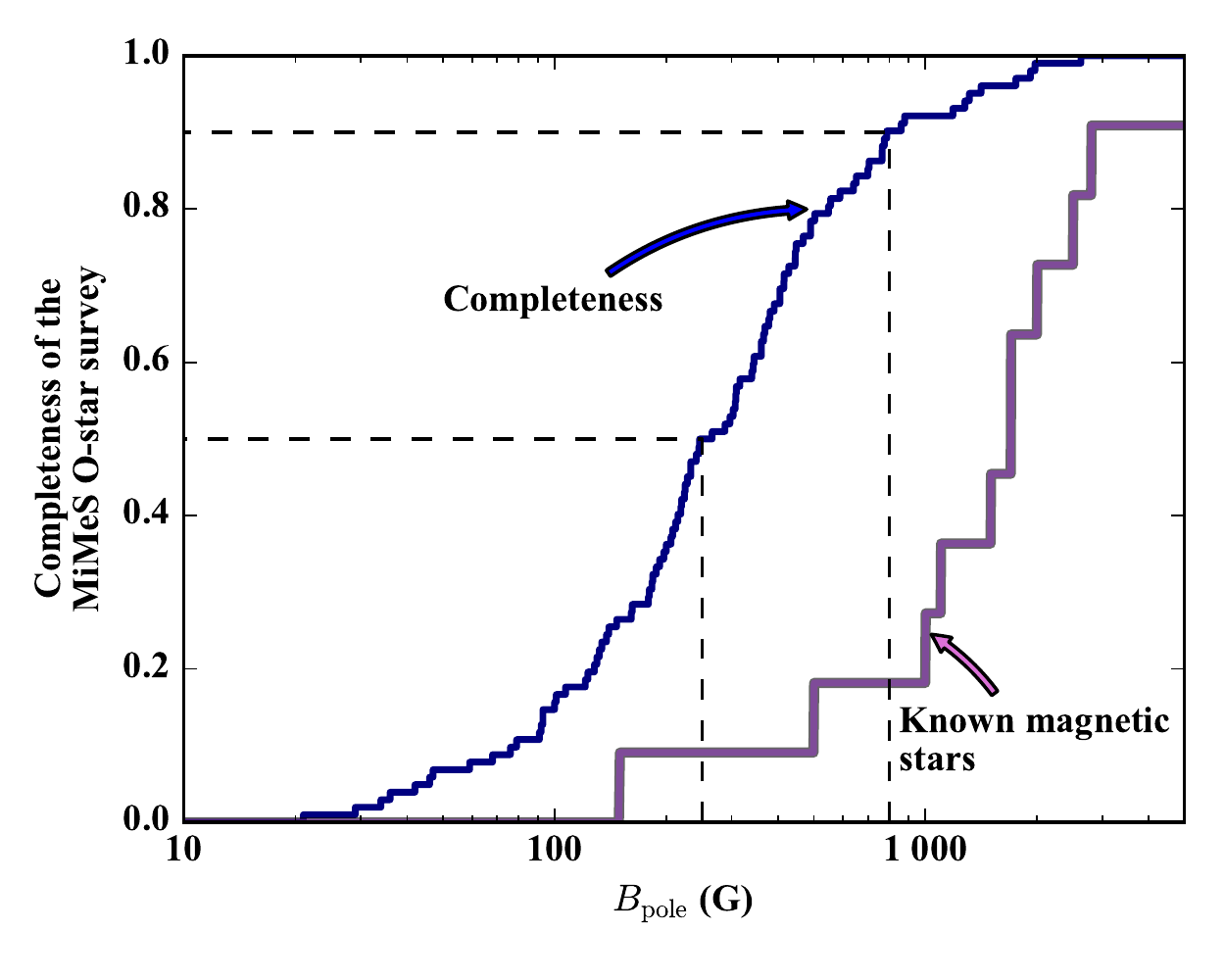}
	\caption{ \label{fig-complete} Completeness of the MiMeS survey of O-type stars derived from the cumulative distribution of the upper limits to the 95.4 percent credible regions for $B_\mathrm{pole}$. The cumulative distribution of field strengths for known magnetic O-type stars is shown as a reference. The dashed lines illustrates the 50 percent and 90 percent completeness levels.  }
\end{figure}

In Fig.\,\ref{fig-complete} we present this cumulative distribution, as a proxy for the completeness of the MiMeS survey of O-type stars with respect to magnetic field strength. The cumulative distribution of field strengths for the known magnetic O-type stars is also shown as a reference. 
The survey is nearly complete for dipolar magnetic field at the kilogauss level (the top dashed line indicate 90 percent completeness), and 50 percent complete at the 250\,G level (lower dashed line).

%----------
%----------

%###################
%###################

\section{Discussion}
\label{sec:discussion}

In this section, we discuss two applications of our sample of upper limits: (i) evaluating the role of observational biases in the observed deficit of old magnetic stars, and (ii) searching for evidence of a magnetic desert in O-type stars. 

%###################
\subsection{Observational bias in the detection of old magnetic stars}

As discussed in \S\ref{sec:intro}, \citetalias{2016A&A...592A..84F} found evidence suggesting a deficit of old magnetic stars by comparing the ages of known magnetic OB stars with those of a magnitude-limited control sample of OB stars. 

One possible explanation of this result is that the sample of known magnetic stars is somewhat observationally biased. 
The estimation of magnetic field detectability performed by \citetalias{2016A&A...592A..84F} indicated that most of the known magnetic OB stars would still be detectable on the TAMS with present-day instrumentation under the magnetic flux conservation hypothesis.
We here test whether this result holds using our sample of O-type stars for which magnetic detection limits are known, by determining (i) if stars near the TAMS have been studied to detect magnetic fields, and (ii) {whether these stars have been observed with a magnetic detection threshold sufficient to allow for detection of the expected surface field strength under the hypothesis of magnetic flux conservation}. 

\subsubsection{Evolutionary status of the O-stars observed by MiMeS}

%----------
\begin{figure}
	\includegraphics[width=0.5\textwidth]{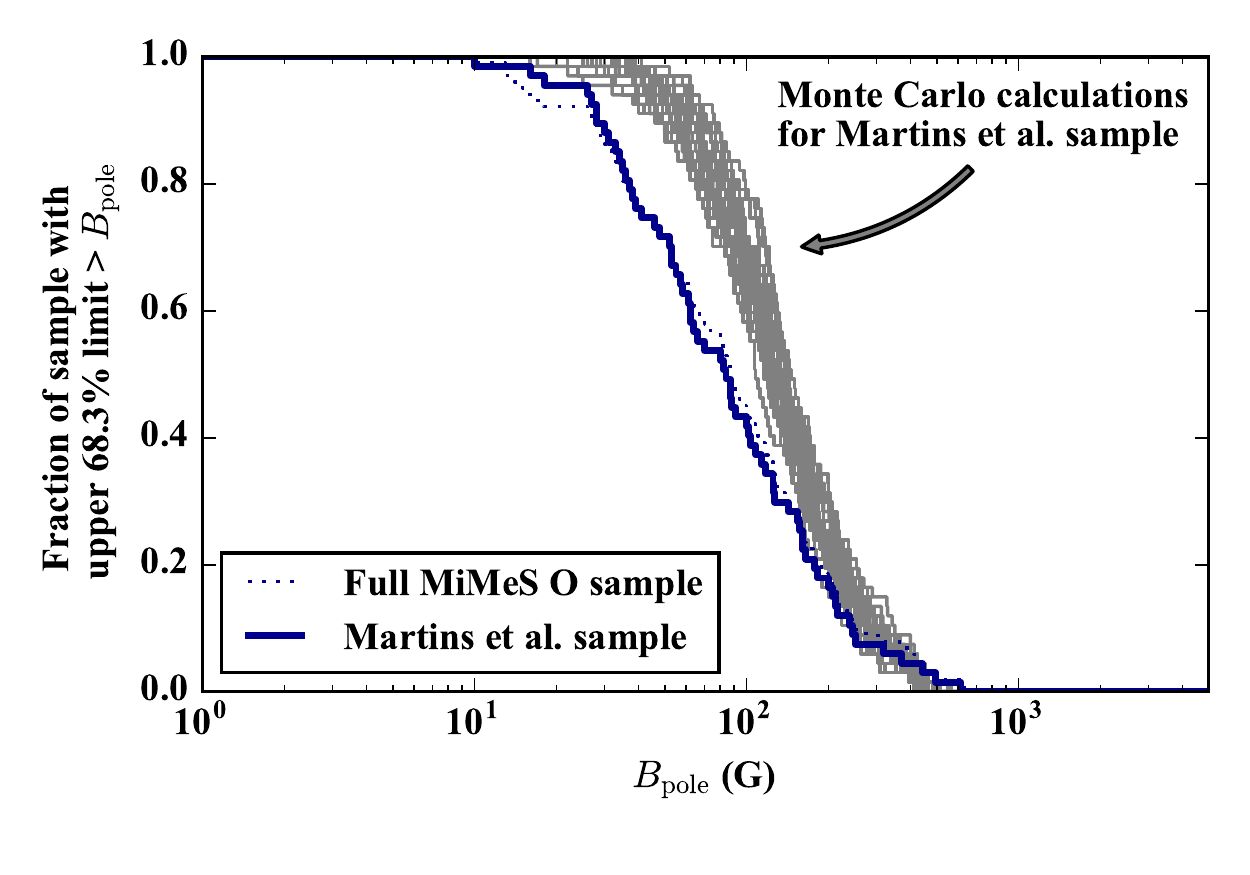}
	\caption{ \label{fig-martins} Cumulative histograms of the fraction of the sample with an upper limit to the 68.3 percent credible region \textit{larger} than a certain dipole strength value $B_\mathrm{pole}$, for the Stokes V profiles. 
	The histogram for the full MiMeS O-star sample (thin dotted blue) is compared to the sample of stars in the \citetalias{2015A&A...575A..34M} (thick solid blue). 
	The Monte Carlo calculations with a dipolar field of 100\,G, restricted to the stars in the \citetalias{2015A&A...575A..34M} sample, are shown in grey.}
\end{figure}
%----------

 In order to derive evolutionary ages in the same fashion as \citetalias{2016A&A...592A..84F}, we select stars in our sample whose physical parameter have already been homogeneously determined by \citet[][hereafter \citetalias{2015A&A...575A..34M}]{2015A&A...575A..34M} using \textsc{cmfgen}  \citep{1998ApJ...496..407H} to model the same MiMeS spectra \citep{2017MNRAS.465.2432G} that are used here. 

 Table \ref{tab:age}, which contains the \textsc{Bonnsai} results described below, lists all the stars in the MiMeS O-type star sample in the same order as in Table \ref{tab:stars}. Empty entries indicate stars not included in the analysis of \citetalias{2015A&A...575A..34M} (for reasons described below).

 \citetalias{2015A&A...575A..34M} excluded known double-lined spectroscopic binaries from their sample. Of the 20 binaries (and their 11 associated O-star companions) included in our O-type star upper limit analysis, only 4 stars remain. This sub-sample of 67 stars is thus biased against those systems. 
 Furthermore, 8 seemingly single stars were only included in the final MiMeS O-type star sample after the parameter modelling of \citetalias{2015A&A...575A..34M} was performed. While these stars were therefore also excluded from the following analysis, they are not representative of a particular class or property, and thus likely do not introduce any obvious bias.

Fig.\,\ref{fig-martins} shows that the upper limits distribution of the \citetalias{2015A&A...575A..34M} sample (thick histogram) are similar to that of the full MiMeS O-star sample (thin histogram). The grey histograms show the upper limit distributions for the Monte Carlo trials with 100\,G injected in all stars, restricted to the stars in the \citetalias{2015A&A...575A..34M} sample.  Compared to the Monte Carlo trials for the full O-type star sample in Fig.\,\ref{fig-upper-MC}, this subset has similar upper limits.

We use the Bayesian code \textsc{bonnsai}\footnote{The \textsc{bonnsai} web-service is available at \url{http://www.astro.uni-bonn.de/stars/bonnsai}.} \citep{2014A&A...570A..66S,2017A&A...598A..60S} to infer the fractional main-sequence ages $\tau$ and other stellar parameters of the \citetalias{2015A&A...575A..34M} sample. To this end, we match the observed surface gravity $\log g$, effective temperatures $T_\mathrm{eff}$ and projected rotational velocities $v\sin i$ against the stellar models of \citet{2011A&A...530A.115B} of solar metallicity. For the prior distributions, we use a Salpeter-like mass function for initial masses \citep{1955ApJ...121..161S}, a uniform distribution for stellar ages, the observed Gaussian distribution of rotational velocities of Milky Way stars of \cite{2008A&A...479..541H} for $v\sin i$ and that all rotation axes are randomly oriented in space. We also checked for differences in the inferred stellar parameters when taking correlations of $\log g$ and $T_\mathrm{eff}$ into account as has been done by \citetalias{2016A&A...592A..84F} for their stellar sample. The resulting $\tau$ distributions were unaffected. We here present stellar parameters without taking correlations into account as the exact correlations of $\log g$ and $T_\mathrm{eff}$ for each star are unknown in our case because of a different technique that has been used to determine the atmospheric parameters \citepalias{2015A&A...575A..34M}.
 The derived fractional main sequence age used in the discussion below are listed in Table \ref{tab:age}.

%----------
\begin{figure}
	\includegraphics[width=0.48\textwidth]{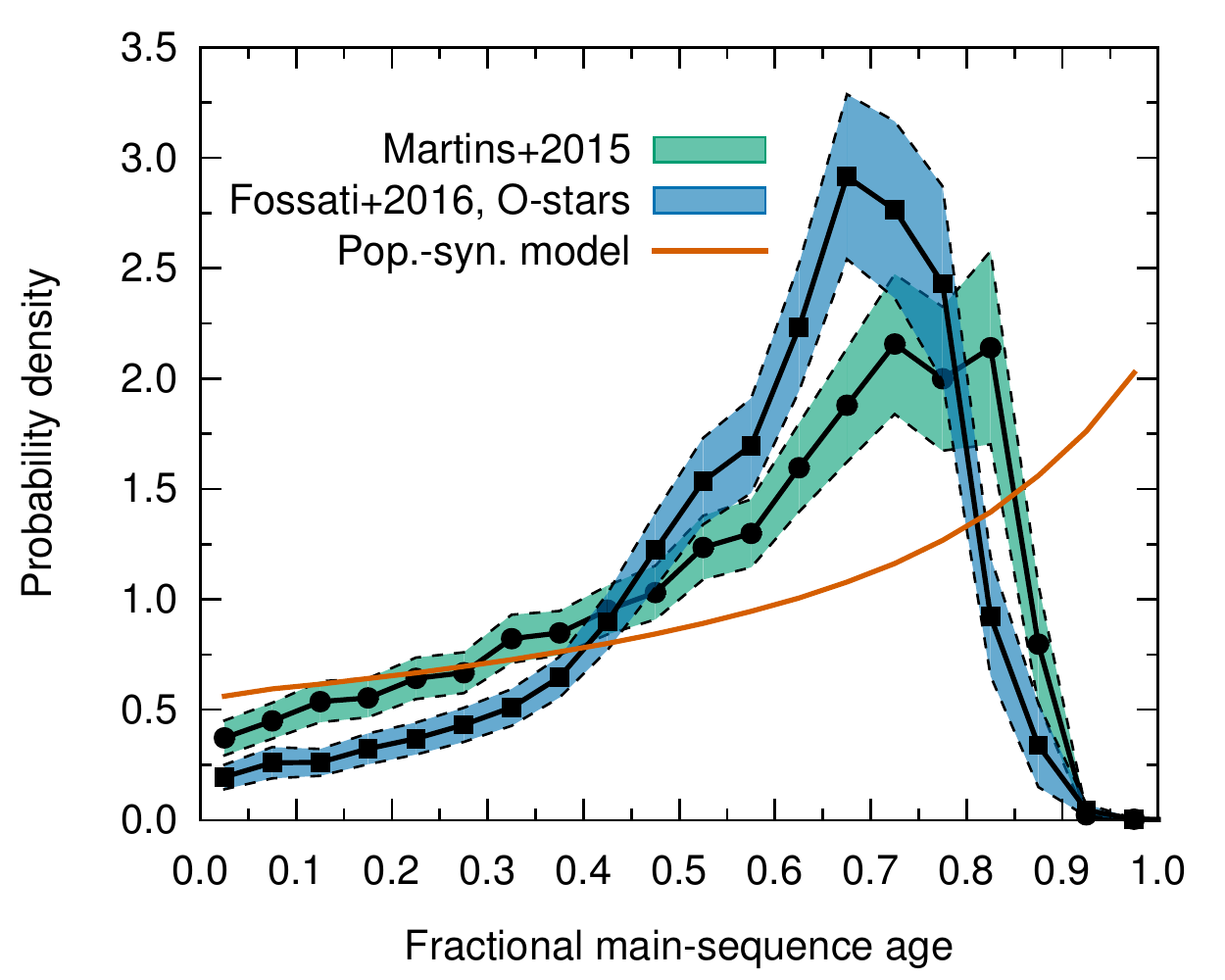}
	\caption{ \label{fig-comparison-tau-pdfs-o-stars-only} Fractional MS $\tau$-distributions of (i) the subsample of the MiMeS O-type star sample analysed by \citetalias{2015A&A...575A..34M} (green) and (ii) the magnitude limited-sample of OB stars presented by \citetalias{2016A&A...592A..84F} but here with only the O-type stars included. The shaded regions indicate bootstrapped 1$\sigma$ estimates of the statistical significance of the variability in the $\tau$-distributions. The solid red line shows the expected $\tau$-distribution from a synthetic population drawn from a magnitude-limited sample \citepalias[see discussion in][]{2016A&A...592A..84F}.  }
\end{figure}
%----------

 Fig.\,\ref{fig-comparison-tau-pdfs-o-stars-only} shows the histogram of the fractional main sequence age ($\tau$) constructed from the sum of each star's posterior probability density, marginalised over all other fitted parameters in the Bayesian framework. In other words, each star contributes to multiple $\tau$ bins, with a distribution following its posterior probability. 
 The sample of MiMeS O-type stars in the \citetalias{2015A&A...575A..34M} sample (in green) is compared with the magnitude-limited control sample of \citetalias{2016A&A...592A..84F} that we modified to include only the O-type stars\footnote{The resulting distribution is very similar to the distribution of OB stars presented by \citetalias{2016A&A...592A..84F}, with a slight shift of the peak of the distribution towards older ages.} (in blue).  
The two fractional MS age distributions are qualitatively very similar, both peaking at $\sim75$ percent of the main sequence lifetime.

 From this we conclude that the MiMeS survey did indeed observe a significant number of O-type stars that are in the second half of their main sequence lifetime. 
 This therefore implies that the inferred deficit of old magnetic O-type stars {is not caused by a lack of observations of more evolved stars by spectropolarimetric surveys}.

%%%
\subsubsection{Is the MiMeS sample good enough to test models of MS field evolution?}
\label{sec:evol}

%----------
\begin{figure*}
	\includegraphics[width=\textwidth]{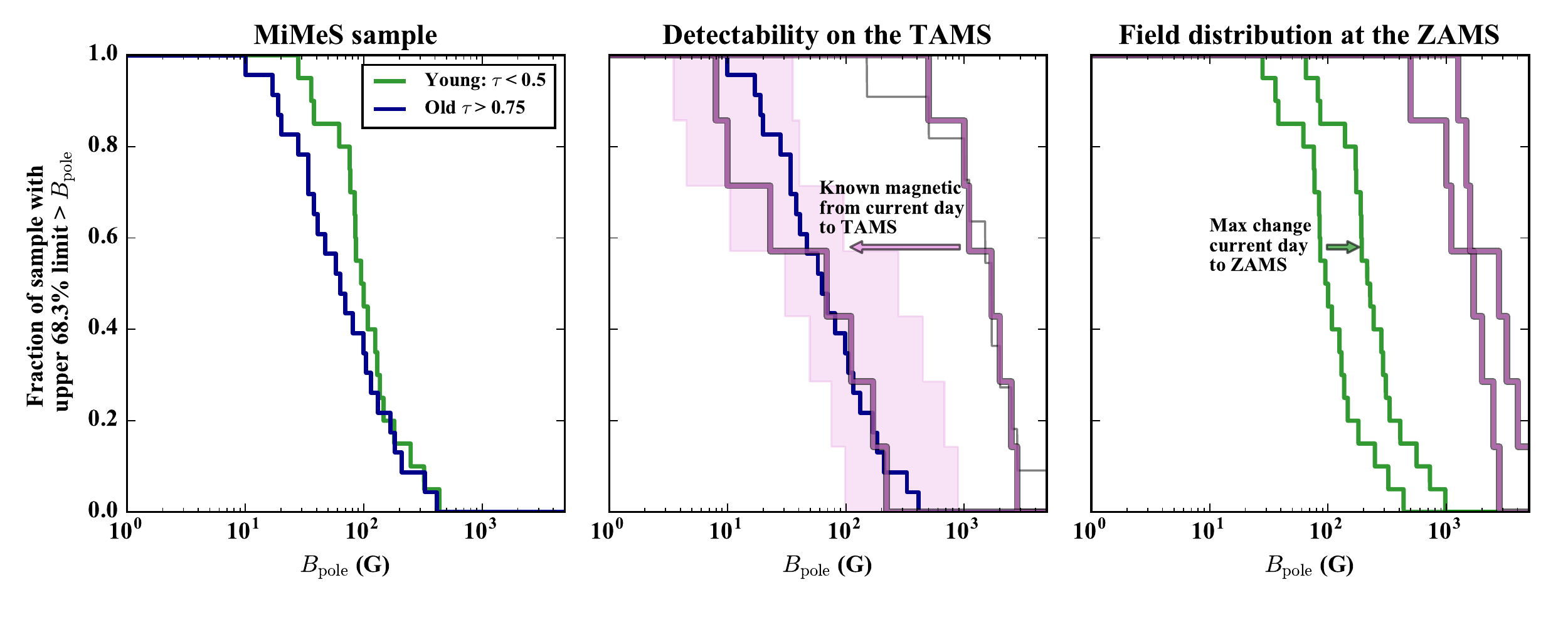}
	\caption{ \label{fig-TAMS} \textit{Left panel}: Cumulative histogram of 68.3 percent Stokes V upper limits for the \citetalias{2015A&A...575A..34M} sample (thick blue histogram of Fig.\,\ref{fig-martins}), separated into a young star sample ($\tau<0.5$; thick green) and an old star sample ($\tau>0.75$; thick blue).  
	\textit{Middle panel}: Comparison of the old sample (thick blue) with the field strength distribution of the known magnetic O-type stars included in the sample of \citetalias{2016A&A...592A..84F} (two thick pink histograms), evolving them from their current age values (right histogram) to their expected field strength at the TAMS (left histogram), assuming magnetic flux conservation. As the sample of \citetalias{2016A&A...592A..84F} only contained magnetic O-type stars with $V<9$, we compare their current day field strength distribution with the field strength distribution of all known magnetic O-type stars we used previously (thin grey histogram). The pink shaded region take into account a systematic uncertainty of 50 percent in the radius change between current age and TAMS.  
	\textit{Right panel}: Comparison between the field upper limits for the young star sample (left-most green histogram) and the current-day magnetic field distribution of known magnetic O-type stars (left-most, thick pink histogram). The right-most green curve shows the maximum field these stars could have on the ZAMS, based on a maximum radius increase by a factor of 1.5.  The right-most pink curve shows the expected ZAMS distribution of field in known magnetic O-type stars based on the \textsc{bonnsai} models of \citetalias{2016A&A...592A..84F}. }
\end{figure*}
%----------

%----------
\begin{figure}
	\includegraphics[width=0.5\textwidth]{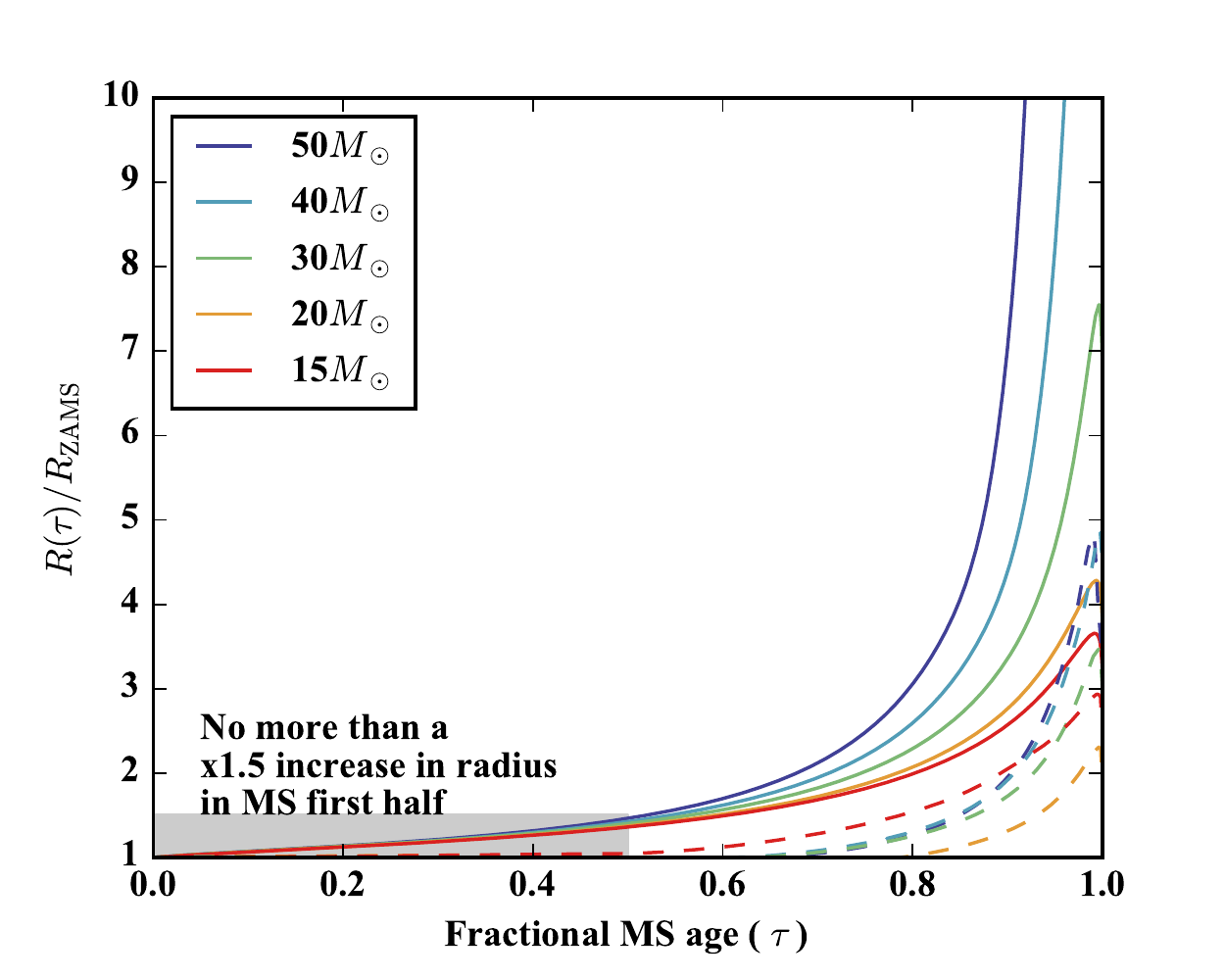}
	\caption{ \label{fig-Brott}  {Variation of the radius as a function of the fractional main sequence age for a selection of models from the \citet{2011A&A...530A.115B} evolution tracks. The initial rotation rates are bracketed between no rotation (in solid curves) and highest available rotation in the model grid, close to critical rotation (in dashed curves).}  }
\end{figure}
%----------

 Now, we seek to determine whether the more evolved stars in the MiMeS O-type star sample have been observed with sufficient precision to detect magnetic fields of the typical strength expected under the magnetic flux conservation hypothesis, i.e. that the surface field strength decreases with time ($t$) as $B(t)/B_o = (R_o/R(t))^2$, where $B_o$ and $R_o$ are the surface field strength and radius at an earlier time.   

Our reasoning is that if we find that these expected field strengths at the TAMS would have led to unambiguous detections within the capabilities of the MiMeS survey, we can rule out that the observational lack of old magnetic O-type stars is caused by a detectability bias.

 Fig.\,\ref{fig-TAMS} (left) shows the distribution of the 68.3 percent upper limits from the MiMeS sample of \citetalias{2015A&A...575A..34M}, separated into two sub-samples according to the peak of their fractional main sequence age probability distribution.  
The ``old'' sample contains the 20 stars with a peak age greater than 75 percent of their main sequence lifetime (thick blue histogram).  
The ``young'' sample contains the 23 stars with a peak age less than 50 percent of the main sequence lifetime (thick green histogram). 
These age thresholds were chosen to separate the two groups as much as possible in fractional age, while still retaining a large enough sample of stars in each group.

 At first glance we can see that older O-type stars in the MiMeS sample have somewhat better magnetic upper limits than younger stars (median of the 68.3 percent upper limits for each sample are 64 and 96\,G, respectively), presumably due to their higher luminosity and slower rotation. 
 However, we also need to assess whether these better upper limits can counterbalance the expected decrease in surface field strength. 

 We do so by performing the same experiment as \citetalias{2016A&A...592A..84F} (their \S\,4.1).
In this experiment, we calculate the expected field strengths that the \textit{known} magnetic O-type stars will have at the TAMS, assuming the surface magnetic flux is conserved from the current age. We then compare this distribution with the upper limit distribution for the old O-type stars in the MiMeS sample. 

 This comparison comes with some caveats:
 
\noindent (i) The middle panel of Fig.\,\ref{fig-TAMS} shows the distribution of field strengths for the 7 known magnetic O-type stars in the sample of \citetalias{2016A&A...592A..84F} (rightmost thick pink histogram). \citetalias{2016A&A...592A..84F} applied a magnitude and luminosity class cut to their samples of magnetic and non-magnetic stars, therefore rejecting 4 known magnetic O-type stars\footnote{NGC 1624-2, CPD-28 2561, Tr16-22, $\zeta$ Ori A.} from the complete distribution of known magnetic O-type stars we presented in Fig.\,\ref{fig-comp-Mag} (reproduced here in Fig.\,\ref{fig-TAMS} with a thin grey line).

\noindent (ii) We thus proceed here with the sample from \citetalias{2016A&A...592A..84F} as we use the change in radius, from current day to TAMS, that they determined with \textsc{bonnsai}, for a direct comparison to their results (relevant values are listed in Table\,\ref{tab:mag}). 
The resulting distribution of expected field strengths at the TAMS for known magnetic O-type stars is shown in Fig.\,\ref{fig-TAMS} (middle panel) as the leftmost thick pink histogram.

\noindent (iii) As we are attempting to determine whether we can rule out a detection bias, we here consider the most optimistic scenario in which stars with magnetic field values greater or equal to the 68.3 percent upper limit are likely to be detected. The 68.3 percent upper limit distribution for the old O-type stars in the MiMeS sample is reproduced from the left panel to the middle panel (blue thick histogram). This distribution overlaps with the distribution of expected field strengths at the TAMS for known magnetic O-type stars. 

\noindent (iv) We now demonstrate how, in this optimistic scenario, the detectability of known magnetic stars at the TAMS also hinges on our level of confidence in the uncertainty of the estimated change of radius experienced by the known magnetic stars on their way to the TAMS.

\begin{table}
\caption{ \label{tab:mag} List of the magnetic O-type stars considered in the study of \citetalias{2016A&A...592A..84F}, with the field strength values as well as the current-day, ZAMS, and TAMS radii determined by \citetalias{2016A&A...592A..84F} and used in \S\ref{sec:evol}. }
\begin{tabular}{l c c c c c}
\hline
Star & $B_\mathrm{pole}$ & $\tau_\mathrm{MS}$ & $R_\mathrm{current}$ & $R_\mathrm{ZAMS}$ & $R_\mathrm{TAMS}$\\
         & (kG) &                                                         & $R_\odot$ & $R_\odot$ &  $R_\odot$ \\
(1) & (2) & (3) & (4) & (5) & (6) \\
 \hline
HD\,148937&1.0&0.4&9.00&7.41&95.3\\
HD\,37022&1.1&0.1&7.19&6.72&48.9\\
HD\,191612&2.5&0.65&11.6&7.08&69.2\\
HD\,47129&2.8&0.4&6.30&5.25&22.4\\
HD\,108&0.5&0.69&13.17&7.39&93.1\\
HD\,57682&1.7&0.53&7.43&5.81&29.0\\
HD\,54879&2.0&0.48&6.88&5.4&23.6\\
\hline
\end{tabular}
\end{table}

To evolve the observed fields strengths to the TAMS, we require an estimate of the current and TAMS radii of the stars. Fig.\,\ref{fig-Brott} shows the variation of radius as a function of fractional main sequence age for a selection of models from the \citet{2011A&A...530A.115B} evolution tracks. A range of initial masses, spanning the range of masses for known magnetic O-type stars, are colored according to the legend, and the rotation rates are bracketed between no rotation (in solid curves) and highest available rotation in the model grid, close to critical rotation (in dashed curves). 

 As can be seen from Fig.\,\ref{fig-Brott}, the change in radius in the second half of the main sequence is very model dependent within a single set of evolutionary tracks.  While most magnetic stars are currently slow rotators, it is unclear if they began the MS as such, or if they started the MS close to critical rotation and experienced the expected rapid magnetic braking \citep{2009MNRAS.392.1022U}. According to the models of \citet{2011A&A...525L..11M}, these two scenarios would result in different interior structures. It is thus unclear which of the sets of non-magnetic models, if any, would be the most appropriate to describe magnetic evolution.
 
 It may also be that the evolutionary tracks for stars with a strong fossil magnetic field are fundamentally different from that of non-magnetic stars due to the field interactions with the convection and angular momentum transport \citep[e.g.][]{2011A&A...525L..11M,2017MNRAS.466.1052P,2018CoSka..48..124K}.  This could mean that (a) the radius evolution is different than that of non-magnetic stars (i.e. the stars stay more/less compact) and/or (b) the real age and main sequence lifetime of the magnetic star might be different from that implied by the non-magnetic evolutionary tracks.  

Therefore, a systematic uncertainty in the ratio of radii of at least 50 percent would certainly not be surprising. The shaded pink region in the middle panel of Fig.\,\ref{fig-TAMS} represents a systematic uncertainty of 50 percent in the radius increase for each known magnetic O-type star. 
Comparing the expected TAMS field strengths of known magnetic stars with the upper limits for stars near the end of their main sequence lifetime, it is clear that not all stars would have been detected by the MiMeS survey, even if their TAMS radius is 50 percent smaller than predicted. 

Specific predictions for the rate at which older magnetic stars would have been detected by the MiMeS survey are therefore very dependent on the details of the evolution models used to predict the radius increase. 
This uncertainty in the radius increase would not have been too critical if the detection limits were, say, 10 times better, so that the thick blue histogram in the middle panel of Fig.\,\ref{fig-TAMS} was located completely to the left of the pink shaded region. 

We remind the reader that our comparison distribution corresponds to 68.3 percent confidence, i.e. the optimistic assumption about detectability. 
While adopting e.g; the 95.4 percent confidence distribution (shifting the thick blue histogram to the right) relaxes the tension between the two distributions, our conclusions do not change: typical magnetic O stars observed at the end of their MS evolution would have been difficult to detect with the precision achieved in the MiMeS survey.

 \textbf{We thus conclude that better upper limits, by at least a factor of 10, would be necessary for a clear and model-free exclusion of an observational bias to explain the apparent lack of old stars in the population of known magnetic O-type stars.}

%###################
\subsection{Magnetic desert}

 We now use our sample of upper limits to evaluate whether there is an extension of the ``magnetic desert'' phenomenon for O-type stars. 
 As discussed in \S\ref{sec:intro-desert}, this so-called magnetic desert designates a feature in the distribution of field strengths of magnetic A and B stars, such that there is only a handful of known AB stars with detected fields of $\sim$100G, even though the magnetic sensitivity for this type of stars is often at the 1-10\,G level. 

This magnetic desert in AB stars has an important diagnostic power for both stellar evolution models and for scenario for the origin of magnetism. 
{Indeed, predicted ``Initial B-field Function'' (IBF) combined with subsequent stellar evolution must result in an evolution of the surface magnetic field strength with time that prevents a long evolution phase at which the field would be below the magnetic desert cutoff but above the detection limit. }
Also, this magnetic desert implies that there are very few young magnetic AB stars born (or in the merger scenario have field generated) with a $ \lessapprox 100$\,G field, pointing toward a bimodality of the IBF.

 In order to test whether the IBF of magnetic O-type stars might be bimodal let us first consider the distribution of upper limits for the young sub-sample in the MiMeS survey we presented in the previous section as the green histogram in the left panel of Fig.\,\ref{fig-TAMS}. This distribution is reproduced in the rightmost panel of Fig.\,\ref{fig-TAMS}. 

 Once again assuming that surface magnetic flux is conserved on stellar evolution timescales, we can make a conservative estimate of the change of radius experienced by these stars since they were on the ZAMS by finding the maximum change in radius for an O-type star during the first half of the main sequence (we remind the reader that this corresponds to the maximum peak age of the young sample) for the set of evolution tracks already presented in Fig.\,\ref{fig-Brott}. For all models, the maximum increase in radius is of a factor 1.5. 

 Let us now imagine that all the stars in this young sample possess an undetected, current-day field with a dipolar value less than or equal to its upper limit. 
If each star could only have increased its radius by a factor of 1.5 at most since the ZAMS, then the ZAMS strength of this imaginary, undetected magnetic field could not have exceeded the current-day upper limit by more than a factor of 2.25 (or a shift of 0.35\,dex in $\log(B_\mathrm{pole})$).  
This ZAMS upper limit distribution is represented by the rightmost green histogram in Fig.\,\ref{fig-TAMS}. 

 Comparing the lack of overlap between this ZAMS upper limit distribution and 
(i) the current-day distribution of field in known magnetic O-type stars (leftmost thick pink histogram),
(ii) the expected ZAMS distribution of fields in known magnetic O-type stars based on the \textsc{bonnsai} models of \citetalias{2016A&A...592A..84F} (rightmost thick pink histogram),
 there is a good indication that the ZAMS IBF is bimodal: that young stars are expected to either have weak/absent magnetic fields, or strong magnetic fields comparable to those of the known magnetic O-type stars. 
In other words, if a \textit{large} fraction of young stars had magnetic fields with a strength of a few hundreds of gauss, evidence for these fields would be found in the MiMeS sample, as described in \$\,\ref{sec:pop}.

%###################
%###################

\section{Conclusions}
\label{sec:conclusion}

In this paper, we perform a statistical analysis of the MiMeS Survey of O-type stars, to explore the dipolar field strengths allowed by the polarisation spectra that lack any magnetic detection. We directly model the Stokes V profiles using Bayesian inference and adopting a dipolar topology.

We conclude that apart from the three magnetic candidate stars (HD\,36486, HD\,162978, HD\,199579) identified by \citet{2017MNRAS.465.2432G} and the one additional magnetic candidate (HD\,37468A) identified here, there is no strong evidence for a magnetic signal consistent with a dipolar field in any other individual star in our sample. We confirm that the 6 probable spurious detections identified by \citet{2017MNRAS.465.2432G} are indeed not compatible with a Stokes V signature arising from a dipolar magnetic field. The addition of one additional magnetic candidate does not change the magnetic field incidence of $7\pm3$ percent reported by \citet{2017MNRAS.465.2432G} within the error bar. 

We determine that if all the non-detected stars were hosting a magnetic field just below their detection limit, the distribution in strength of these undetected fields would be different from the distribution in strength of the known magnetic O-type stars. 

Considering all of the target stars as a sample, we find a good correlation between the statistics obtained from the Stokes V profiles and the null profiles (which are nominally flat and expected to potential display only instrumental polarisation or non-magnetic variability, if any). Using a Monte Carlo calculation, we also conclude that if all the stars in our sample were hosting a 100\,G dipolar magnetic field,  the direct detection rate would be rather low, only slightly higher than the observed bulk incidence of magnetism in massive stars. 
From this, we assess that the survey is nearly complete with respect to the detection of a lone magnetic star for dipolar field at the kilogauss level, and 50 percent complete at the 250\,G level.
However from a survey sample point of view, the scenario above (all stars hosting a magnetic field) would have resulted in statistically different odds ratio and credible region upper limit distributions obtained from Stokes V and from the null profiles.

We address two questions concerning the distribution of magnetic fields in O-type stars: (i) whether the ``magnetic desert'' phenomenon extends to the O-type stars, and (ii) whether the MiMeS sample of O-type star was able to systematically detect magnetic fields in old main sequence O-type stars, and thereby constrain models of the evolution of their surface magnetism. 

We find a good indication that the Initial B-field Function is bimodal -- young O-type stars are expected to either have weak/absent magnetic fields, or strong magnetic fields. 

We also find that better upper limits -- better by at least a factor of 10 -- would be necessary to affirm that the known magnetic O-type stars would have been unambiguously detected by the MiMeS Survey had they been on the TAMS. In other words, the current detection limits for the old stars in the MiMeS sample, combined with large uncertainties in the increase of the stellar radii over time in stellar evolution models, do not allow us to rule out a detection bias to explain the apparent lack of old magnetic O-type stars reported by \citet{2016A&A...592A..84F}.

\section*{Acknowledgments}

Based on MiMeS LP and archival spectropolarimetric observations obtained at the CFHT which is operated by the National Research Council of Canada, the Institut National des Sciences de l'Univers (INSU) of the Centre National de la Recherche Scientifique (CNRS) of France, and the University of Hawaii; 
on MiMeS LP and archival observations obtained using the Narval spectropolarimeter at the Observatoire du Pic du Midi (France), which is operated by CNRS/INSU and the University of Toulouse; 
and on MiMeS LP observations acquired using HARPSpol on the ESO 3.6 m telescope at La Silla Observatory, Program ID 187.D-0917.
This research has made extensive use of the SIMBAD database, operated at CDS, Strasbourg, France. 
We acknowledge the CADC.

VP acknowledges support from the Fonds de Recherche Nature et Technology, the ESO Scientific Visitor Programme, This material is based upon work supported by the National Science Foundation under Grant No. 1747658.

ADU gratefully acknowledges the support of the Natural Science and Engineering Research Council of Canada (NSERC).

The MiMeS collaboration acknowledges financial support from the Programme National de Physique Stellaire (PNPS) of INSU/CNRS.

The authors would like to thank Dr. J. Grunhut for his contributions to the MiMeS project that made the work presented in this paper possible. 

\bibliographystyle{mn2e_fix2}
\bibliography{database}

%\processdelayedfloats

\newpage
\onecolumn
		{\setlength\LTcapwidth{\linewidth}
\begin{longtable}{ l  l c  c  c  c  c  r  r  r  r  r  r  }
\caption{\label{tab:stars}Results from our Bayesian analysis. The columns report: (1) a reference number with binarity status, (2) the name of the star with (3) its spectral type reproduced from \citetalias{2017MNRAS.465.2432G}, (4) the number of nightly-averaged observations, (5) the rotational broadening adopted in our models, the odds ratio $\log( M_{B_p=0} / M_{B_p} )$ obtained from (6) the Stokes V profile and (7) from the null profiles, (8) the upper limit to the 68.3 percent credible region obtained from Stokes V, (9) same as previous column for the 95.4 percent credible region, (10-11) same as the two previous columns for the null profiles, the mode of the probability density function marginalized for $B_p$ obtained (12) from Stokes V and (13) from the null profiles. The bold red entries highlight odd ratios $\log( M_{B_p=0} / M_{B_p} )<-0.5$ for easy reference.}\\
\hline
\mcc{ID} & \mcc{Name} & \mcc{Spec. type} & \mcc{Nobs} & \mcc{$v\sin i$} & \mcc{Odds V} & \mcc{Odds N} & \mcc{68 V} & \mcc{95 V} & \mcc{68 N} & \mcc{95 N} & \mcc{Mode V} & \mcc{Mode N} \\
\mcc{} & \mcc{} & \mcc{}  & \mcc{} & \mcc{km\,s$^{-1}$} & \mcc{} & \mcc{} & \mcc{G} & \mcc{G} & \mcc{G} & \mcc{G} & \mcc{G} & \mcc{G} \\
\mcc{(1)} & \mcc{(2)} & \mcc{(3)} & \mcc{(4)} & \mcc{(5)} & \mcc{(6)} & \mcc{(7)} & \mcc{(8)} & \mcc{(9)} & \mcc{(10)} & \mcc{(11)} & \mcc{(12)} & \mcc{(13)}\\
\hline
\endfirsthead
\caption{Continued}\\
\hline
\mcc{ID} & \mcc{Name} & \mcc{Spec. type} & \mcc{Nobs} & \mcc{$v\sin i$} & \mcc{Odds V} & \mcc{Odds N} & \mcc{68 V} & \mcc{95 V} & \mcc{68 N} & \mcc{95 N} & \mcc{Mode V} & \mcc{Mode N} \\
\mcc{} & \mcc{} & \mcc{} & \mcc{} & \mcc{km\,s$^{-1}$} & \mcc{} & \mcc{} & \mcc{G} & \mcc{G} & \mcc{G} & \mcc{G} & \mcc{G} & \mcc{G} \\
\mcc{(1)} & \mcc{(2)} & \mcc{(3)} & \mcc{(4)} & \mcc{(5)} & \mcc{(6)} & \mcc{(7)} & \mcc{(8)} & \mcc{(9)} & \mcc{(10)} & \mcc{(11)} & \mcc{(12)} & \mcc{(13)} \\
\hline
\endhead
\hline
\multicolumn{13}{r}{Following on the next page} \\
\endfoot
\hline
\multicolumn{13}{l}{$^\dagger$ Stars identified as binaries in \citetalias{2017MNRAS.465.2432G} but for which a SB2 treatment }\\
\multicolumn{13}{l}{was unnecessary (SB1, large luminosity ratio, etc) or unfeasible (see notes in Appendix \ref{sec:apSB2}). }\\
\endlastfoot
1	SB2	&	HD\,	1337	A	&O9.5\,II(n)&	1	&	118	&	0.02	&	0.41	&	212	&549	&63	&225	&132	&	0	\\
2		&		HD\,	1337	B	&&	1	&	23	&	0.28	&	0.32	&	275	&1187	&232	&931	&0	&	0	\\
3		&	HD\,	13745		&O9.7\,II(n)&	3	&	170	&	0.2	&	0.46	&	177	&491	&105	&314	&0	&	0	\\
4		&	HD\,	14633		&ON8.5\,V&	1	&	120	&	0.5	&	0.19	&	58	&209	&118	&372	&0	&	0	\\
5	SB3	&	HD\,	17505	Aa1	&O6.5\,IIIn((f))&	1	&	8	&	0.27	&	0.28	&	423	&1976	&358	&1727	&0	&	0	\\
6		&	HD\,	17505		Ab	&&	1	&	14	&	0.3	&	0.22	&	435	&1921	&585	&2922	&0	&	0	\\
7		&	HD\,	17505		Aa2	&&	1	&	1	&	0.16	&	0.26	&	614	&2628	&413	&2075	&0	&	0	\\
8		&	HD\,	24431		&O9\,III&	1	&	60	&	0.29	&	0.18	&	48	&162	&70	&218	&0	&	0	\\
9		&	HD\,	24534		&O9.5npe&	1	&	200	&	\red{-0.73}	&	0.42	&	632	&1414	&123	&443	&387	&	0	\\
10		&	HD\,	24912		& O7.5\,III(n)((f))&	13	&	201	&	0.64	&	\red{-1.27}	&	13	&37	&50	&125	&0	&	27	\\
11		&	HD\,	30614		&O9\,Ia&	5	&	96	&	0.64	&	0.54	&	16	&47	&19	&56	&0	&	0	\\
12		&	HD\,	34078		&O9.5\,V&	4	&	16	&	0.67	&	0.9	&	29	&69	&16	&49	&0	&	0	\\
13		&	HD\,	34656		&O7.5\,II(f)&	1	&	70	&	0.37	&	0.39	&	52	&185	&41	&148	&0	&	0	\\
14		&	HD\,	35619		&O7.5\,V&	1	&	45	&	0.18	&	0.31	&	108	&341	&72	&248	&0	&	0	\\
15	SB2	&	HD\,	35921		&O9.5\,II&	1	&	205	&	0.42	&	0.42	&	103	&370	&125	&448	&0	&	0	\\
16	$^\dagger$	&	HD\,	36486		&O9.5\,IINwk&	2	&	121	&	\red{-2.46}	&	0.64	&	94	&242	&21	&70	&56	&	0	\\
17		&	HD\,	36512		&O9.7\,V&	1	&	18	&	0.48	&	0.17	&	33	&122	&86	&255	&0	&	0	\\
18		&	HD\,	36861		&O8\,III((f))&	8	&	61	&	0.61	&	0.65	&	14	&35	&12	&35	&0	&	0	\\
19		&	HD\,	36879		&O7\,V(n)((f))&	2	&	185	&	0.52	&	\red{-0.98}	&	55	&180	&259	&556	&0	&	154	\\
20	SB2	&	HD\,	37041		&O9.5\,Ivp&	3	&	123	&	-0.03	&	0.58	&	84	&193	&31	&93	&0	&	0	\\
21	SB2	&	HD\,	37043		&O9\,III\,var&	1	&	78	&	0.02	&	0.35	&	80	&220	&30	&104	&0	&	0	\\
22	SB2	&	HD\,	37366		&O9.5\,IV&	1	&	2	&	0.25	&	0.32	&	40	&207	&29	&167	&0	&	0	\\
23	SB2	&	HD\,	37468		&O9.7\,III&	1	&	118	&	\red{-1.62}	&	0.22	&	182	&406	&56	&180	&112	&	0	\\
24		&	HD\,	38666		&O9.5\,V&	1	&	114	&	0.37	&	-0.35	&	38	&135	&100	&273	&0	&	0	\\
25		&	HD\,	42088		&O6\,V((f)z&	1	&	37	&	0.32	&	0.38	&	113	&382	&85	&303	&0	&	0	\\
26		&	HD\,	46056		&O8\,Vn &	1	&	344	&	0.31	&	0.37	&	371	&1279	&294	&997	&0	&	0	\\
27	SB2	&	HD\,	46106		&O9.7\,II-III&	1	&	86	&	0.44	&	0.37	&	88	&317	&124	&429	&0	&	0	\\
28	SB2	&	HD\,	46149		& O8.5\,V&	3	&	40	&	0.38	&	0.55	&	70	&309	&38	&114	&0	&	0	\\
29		&	HD\,	46150		&O5\,V((f))z&	3	&	86	&	\red{-0.80}	&	-0.05	&	160	&298	&108	&276	&0	&	0	\\
30		&	HD\,	46202		&O9.5\,V&	2	&	20	&	0.43	&	0.48	&	31	&94	&28	&92	&0	&	0	\\
31		&	HD\,	46223		&O4\,V((f))&	2	&	67	&	0.23	&	0.41	&	126	&344	&77	&248	&0	&	0	\\
32		&	HD\,	46485		&O7\,Vn&	3	&	312	&	0.51	&	0.38	&	127	&379	&192	&530	&0	&	0	\\
33		&	HD\,	46966		&O8.5\,IV&	1	&	50	&	0.43	&	0.39	&	39	&139	&39	&137	&0	&	0	\\
34		&	HD\,	47432		&O9.7\,Ib&	1	&	90	&	0.42	&	0.28	&	30	&108	&47	&161	&0	&	0	\\
35	$^\dagger$	&	HD\,	47839		&O7\,V((f))\,var&	8	&	50	&	0.68	&	0.49	&	15	&43	&27	&68	&0	&	0	\\
36	SB2	&	HD\,	48099	A	&O6.5\,V(n)((f))&	2	&	91	&	0.27	&	0.54	&	82	&226	&41	&141	&0	&	0	\\
37		&	HD\,	48099		B	&&	2	&	51	&	0.63	&	0.58	&	52	&184	&58	&190	&0	&	0	\\
38	SB2	&	HD\,	54662	A	&O7\,V((f))z\,var?&	1	&	31	&	0.34	&	0.32	&	53	&181	&57	&191	&0	&	0	\\
39		&	HD\,	54662		B	&&	1	&	148	&	0.37	&	0.24	&	87	&304	&137	&423	&0	&	0	\\
40		&	HD\,	55879		&O9.7\,III&	1	&	36	&	0.35	&	0.36	&	16	&60	&17	&60	&0	&	0	\\
41		&	HD\,	66788		&O8\,V&	2	&	24	&	0.3	&	0.37	&	200	&556	&147	&460	&0	&	0	\\
42		&	HD\,	66811		&O4\,If &	2	&	185	&	0.46	&	\red{-0.61}	&	32	&101	&140	&322	&0	&	0	\\
43		&	HD\,	69106		&O9.7\,In&	1	&	312	&	0.41	&	0.38	&	100	&362	&115	&415	&0	&	0	\\
44		&	HD\,	93028		&O9\,IV&	1	&	20	&	0.28	&	0.34	&	129	&447	&122	&465	&0	&	0	\\
45	$^\dagger$	&	HD\,	93250		&O4\,IIIfc:&	1	&	50	&	0.15	&	0.15	&	496	&1755	&435	&1442	&0	&	0	\\
46		&	HD\,	149038		&O9.7\,Iab&	1	&	66	&	0.3	&	0.3	&	28	&94	&28	&96	&0	&	0	\\
47		&	HD\,	149757		& O9.5\,Ivnn&	45	&	352	&	0.77	&	0.6	&	47	&129	&76	&180	&0	&	0	\\
48		&	HD\,	151804		&O8\,Iaf&	1	&	78	&	0.36	&	0.18	&	57	&201	&81	&255	&0	&	0	\\
49		&	HD\,	152233		& O6\,III(f)&	1	&	75	&	0.22	&	0.36	&	120	&367	&74	&262	&0	&	0	\\
50		&	HD\,	152247		&O9.5\,II-III&	1	&	80	&	0.19	&	0.06	&	245	&776	&328	&939	&0	&	0	\\
51		&	HD\,	152249		&OC9\,Iab&	1	&	80	&	0.39	&	0.26	&	36	&131	&58	&187	&0	&	0	\\
52		&	HD\,	152408		&O8\,Iafpe&	1	&	78	&	0.29	&	0.26	&	139	&469	&159	&513	&0	&	0	\\
53	SB2	&	HD\,	153426		&O9\,II-III&	2	&	94	&	0.45	&	-0.03	&	62	&189	&181	&403	&0	&	0	\\
54		&	HD\,	153919		&O6\,Iaf&	1	&	135	&	0.07	&	0.35	&	157	&449	&77	&265	&0	&	0	\\
55		&	HD\,	154368		&O9.5\,Iab&	1	&	73	&	0.4	&	0.34	&	28	&102	&34	&118	&0	&	0	\\
56		&	HD\,	154643		&O9.5\,III&	1	&	106	&	0.39	&	0.41	&	41	&148	&38	&135	&0	&	0	\\
57		&	HD\,	155806		&O7.5\,V&	4	&	70	&	0.46	&	0.4	&	35	&92	&36	&99	&0	&	0	\\
58	SB2	&	HD\,	155889		&O9.5\,IV&	1	&	12	&	0.1	&	0.36	&	198	&862	&98	&391	&0	&	0	\\
59		&	HD\,	156154		&O7.5\,Ibf&	1	&	78	&	0.41	&	0.4	&	62	&221	&63	&224	&0	&	0	\\
60		&	HD\,	162978		&O8\,II(f)&	2	&	76	&	\red{-1.90}	&	0.42	&	161	&289	&26	&81	&91	&	0	\\
61		&	HD\,	164492		&O7.5\,Vz&	1	&	32	&	\red{-0.62}	&	0.34	&	442	&880	&84	&306	&274	&	0	\\
62	SB2	&	HD\,	164794	A	&O3.5\,V((f*))&	5	&	87	&	0.49	&	0.1	&	150	&417	&373	&860	&0	&	0	\\
63		&	HD\,	164794		B	&O5-5.5\,V((f))&	5	&	57	&	0.53	&	0.28	&	85	&247	&157	&414	&0	&	0	\\
64	SB2	&	HD\,	165052	A	&O5.5Vz&	1	&	73	&	0.34	&	0.36	&	110	&392	&98	&350	&0	&	0	\\
65		&	HD\,	165052		B	&O8V&	1	&	80	&	0.4	&	0.24	&	97	&346	&185	&579	&0	&	0	\\
66		&	HD\,	167263		&O9.5\,II-IIIn&	1	&	57	&	-0.19	&	0.3	&	251	&589	&70	&253	&153	&	0	\\
67		&	HD\,	167264		&O9.7\,Iab&	9	&	51	&	0.57	&	0.67	&	27	&80	&19	&55	&0	&	0	\\
68	$^\dagger$	&	HD\,	167771		&O7\,III(f)&	1	&	50	&	0.21	&	0.25	&	64	&198	&56	&190	&0	&	0	\\
69		&	HD\,	186980		&O7.5\,III((f))&	1	&	67	&	0.37	&	0.4	&	37	&133	&37	&131	&0	&	0	\\
70		&	HD\,	188001		&O7.5\,Iabf &	1	&	89	&	0.18	&	0.27	&	165	&504	&132	&419	&0	&	0	\\
71		&	HD\,	188209		&O9.5\,Iab&	7	&	69	&	0.57	&	0.57	&	10	&30	&10	&30	&0	&	0	\\
72		&	HD\,	189957		&O9.7\,III&	1	&	84	&	-0.01	&	0.25	&	142	&429	&99	&337	&0	&	0	\\
73	$^\dagger$	&	HD\,	190918		&WN5o+O9\,I&	2	&	107	&	0.28	&	0.39	&	118	&310	&97	&306	&0	&	0	\\
74	SB2	&	HD\,	191201	A	&O9.5\,III&	1	&	25	&	0.25	&	0.24	&	176	&766	&166	&738	&0	&	0	\\
75		&	HD\,	191201		B	&B0\,IV&	1	&	2	&	0.24	&	0.23	&	162	&765	&173	&816	&0	&	0	\\
76		&	HD\,	192281		& O4.5\,Vn(f)&	2	&	270	&	0.06	&	0.32	&	318	&788	&184	&551	&0	&	0	\\
77		&	HD\,	192639		&O7.5\,Iabf&	1	&	96	&	0.43	&	0.12	&	66	&234	&131	&396	&0	&	0	\\
78	SB2	&	HD\,	193322	Aa	&O9\,IV(n)&	1	&	345	&	0.24	&	0.15	&	411	&1315	&439	&1425	&0	&	0	\\
79		&	HD\,	193322		Ab1	&&	1	&	39	&	0.19	&	0.31	&	67	&225	&49	&174	&0	&	0	\\
80	SB2	&	HD\,	193443	A	&O9\,III&	1	&	79	&	0.35	&	0.34	&	72	&246	&74	&255	&0	&	0	\\
81		&	HD\,	193443		B	&&	1	&	74	&	0.34	&	0.4	&	215	&706	&168	&593	&0	&	0	\\
82	$^\dagger$	&	HD\,	199579		&O6.5\,V((f))z&	1	&	35	&	\red{-1.21}	&	0.06	&	154	&416	&95	&344	&0	&	0	\\
83		&	HD\,	201345		& ON9.5\,IV&	1	&	90	&	0.37	&	0.35	&	117	&406	&130	&438	&0	&	0	\\
84		&	HD\,	203064		&O7.5\,IIIn((f))&	4	&	277	&	0.55	&	0.83	&	81	&229	&49	&163	&0	&	0	\\
85	SB2	&	HD\,	204827		&O9.7\,III&	1	&	71	&	-0.11	&	0.37	&	233	&641	&88	&312	&0	&	0	\\
86		&	HD\,	206183		&O9.5\,IV-V&	1	&	9	&	0.38	&	0.43	&	53	&213	&38	&144	&0	&	0	\\
87	$^\dagger$	&	HD\,	206267		&O6.5\,V((f))&	1	&	40	&	0.23	&	0.22	&	132	&491	&127	&460	&0	&	0	\\
88		&	HD\,	207198		&O9\,II&	10	&	69	&	0.6	&	0.47	&	18	&48	&20	&54	&0	&	0	\\
89		&	HD\,	207538		&O9.7\,IV&	1	&	35	&	0.4	&	0.29	&	26	&93	&38	&122	&0	&	0	\\
90	SB2	&	HD\,	209481	A	&O9\,IV(n)\,var&	11	&	135	&	-0.36	&	0.75	&	102	&234	&33	&95	&58	&	0	\\
91		&	HD\,	209481		B	&&	11	&	85	&	0.3	&	0.71	&	217	&446	&77	&217	&0	&	0	\\
92		&	HD\,	209975		&O9\,Ib&	9	&	70	&	\red{-0.67}	&	0.49	&	35	&77	&14	&38	&21	&	0	\\
93		&	HD\,	210809		&O9\,Iab&	1	&	100	&	0.42	&	\red{-1.10}	&	34	&124	&184	&419	&0	&	117	\\
94		&	HD\,	210839		&O6.5\,I(n)fp&	26	&	220	&	0.15	&	\red{-3.44}	&	88	&141	&144	&220	&62	&	116	\\
95		&	HD\,	214680		&O9\,V&	17	&	26	&	-0.07	&	0.53	&	13	&22	&8	&16	&9	&	0	\\
96		&	HD\,	218195		&O8.5\,III&	1	&	61	&	0.41	&	0.27	&	46	&163	&76	&250	&0	&	0	\\
97		&	HD\,	218915		&O9.5\,Iab&	1	&	63	&	0.01	&	0.42	&	91	&267	&34	&122	&0	&	0	\\
98		&	HD\,	227757		&O9.5\,V&	2	&	18	&	0.39	&	0.38	&	238	&701	&323	&1157	&0	&	0	\\
99		&	HD\,	258691		& O9.5\,IV&	1	&	23	&	0.04	&	0.35	&	125	&362	&62	&216	&0	&	0	\\
100		&	HD\,	328856		&O9.7\,II&	1	&	98	&	0.39	&	0.07	&	87	&308	&209	&595	&0	&	0	\\
101		&	BD\,$-$13 4930		&O9.7\,V&	1	&	13	&	0.42	&	0.39	&	61	&216	&76	&280	&0	&	0	\\
102		&	BD\,$+$60 499		&O9.5\,V&	1	&	23	&	0.29	&	0.04	&	207	&652	&249	&724	&0	&	152	\\
\hline
\end{longtable}

\clearpage

\begin{longtable}{ l  l  c  c  c  c  c   }
\caption{\label{tab:age}Results from the \textsc{bonnsai} analysis. Column 1-3 are reproduced from Table \ref{tab:stars}.} \\
\hline
\mcc{ID} & \mcc{Name}& \mcc{Spec. Type} & \mcc{Age} & \mcc{$M_\star$} & \mcc{$R_\star$} & \mcc{MS fractional age}  \\
\mcc{} & \mcc{} & \mcc{} & \mcc{Myr} & \mcc{$M_\odot$} & \mcc{$R_\odot$}  & \mcc{} \\
\mcc{(1)} & \mcc{(2)} & \mcc{(3)} & \mcc{(4)} & \mcc{(5)} & \mcc{(6)} & \mcc{(7)}  \\
\hline
\endfirsthead
\caption{Continued}\\
\hline
\mcc{ID} & \mcc{Name}& \mcc{Spec. Type} & \mcc{Age} & \mcc{$M_\star$} & \mcc{$R_\star$} & \mcc{MS fractional age}  \\
\mcc{} & \mcc{} & \mcc{} & \mcc{Myr} & \mcc{} & \mcc{}  & \mcc{} \\
\mcc{(1)} & \mcc{(2)} & \mcc{(3)} & \mcc{(4)} & \mcc{(5)} & \mcc{(6)} & \mcc{(7)}  \\
\hline
\endhead
\hline
\multicolumn{7}{r}{Following on the next page} \\
\endfoot
\hline
\multicolumn{7}{l}{$^\dagger$ Stars identified as binaries in \citetalias{2017MNRAS.465.2432G} but for which a SB2 treatment }\\
\multicolumn{7}{l}{was unnecessary (SB1, large luminosity ratio, etc) or unfeasible (see notes in Appendix \ref{sec:apSB2}). }\\
\endlastfoot

1	SB2	&	HD\,	1337	A	&O9.5\,II(n)&	&	&	&	\\				
2		&	HD\,	1337	B	&&	&	&	&	\\				
3		&	HD\,	13745		&O9.7\,II(n)&	4.32	&	30	&	19.81	&	0.85	\\
4		&	HD\,	14633		&ON8.5\,V&	3.94	&	22	&	8.16	&	0.66	\\
5	SB2	&	HD\,	17505	Aa1	&O6.5\,IIIn((f))&	&	&	&	\\				
6		&	HD\,	17505	Ab	&&	&	&	&	\\				
7		&	HD\,	17505	Aa2	&&	&	&	&	\\				
8		&	HD\,	24431		&O9\,III&	4.22	&	22	&	8.72	&	0.7	\\
9		&	HD\,	24534		&O9.5npe&	&	&	&	\\				
10		&	HD\,	24912		& O7.5\,III(n)((f))&	3.88	&	25	&	10.93	&	0.71	\\
11		&	HD\,	30614		&O9\,Ia&	4.42	&	26	&	17.38	&	0.84	\\
12		&	HD\,	34078		&O9.5\,V&	4.14	&	18	&	6.2	&	0.48	\\
13		&	HD\,	34656		&O7.5\,II(f)&	3.22	&	26	&	9.4	&	0.66	\\
14		&	HD\,	35619		&O7.5\,V&	3.5	&	22	&	7.29	&	0.55	\\
15	SB2	&	HD\,	35921		&O9.5\,II&	&	&	&	\\				
16	$^\dagger$	&	HD\,	36486		&O9.5\,IINwk&	&	&	&	\\				
17		&	HD\,	36512		&O9.7\,V&	4.36	&	18	&	6.22	&	0.48	\\
18		&	HD\,	36861		&O8\,III((f))&	3.58	&	24	&	9.08	&	0.66	\\
19		&	HD\,	36879		&O7\,V(n)((f))&	3.06	&	28	&	9.64	&	0.62	\\
20	SB2	&	HD\,	37041		&O9.5\,Ivp&	&	&	&	\\				
21	SB2	&	HD\,	37043		&O9\,III\,var&	&	&	&	\\				
22	SB2	&	HD\,	37366		&O9.5\,IV&	&	&	&	\\				
23	SB2	&	HD\,	37468		&O9.7\,III&	&	&	&	\\				
24		&	HD\,	38666		&O9.5\,V&	4.08	&	19	&	6.2	&	0.45	\\
25		&	HD\,	42088		&O6\,V((f)z&	2.44	&	27	&	7.38	&	0.41	\\
26		&	HD\,	46056		&O8\,Vn &	3.9	&	24	&	8.68	&	0.63	\\
27	SB2	&	HD\,	46106		&O9.7\,II-III&	5.84	&	16.6	&	7.26	&	0.68	\\
28	SB2	&	HD\,	46149		& O8.5\,V&	&	&	&	\\				
29		&	HD\,	46150		&O5\,V((f))z&	1.52	&	38	&	8.82	&	0.4	\\
30		&	HD\,	46202		&O9.5\,V&	1.14	&	18	&	5.57	&	0.1	\\
31		&	HD\,	46223		&O4\,V((f))&	1.34	&	41	&	9.23	&	0.37	\\
32		&	HD\,	46485		&O7\,Vn&	3.34	&	26	&	9.4	&	0.59	\\
33		&	HD\,	46966		&O8.5\,IV&	3.5	&	22	&	7.29	&	0.55	\\
34		&	HD\,	47432		&O9.7\,Ib&	4	&	30	&	20.5	&	0.85	\\
35	$^\dagger$	&	HD\,	47839		&O7\,V((f))\,var&	&	&	&	\\				
36	SB2	&	HD\,	48099	A	&O6.5\,V(n)((f))&	&	&	&	\\				
37		&	HD\,	48099	B	&&	&	&	&	\\				
38	SB2	&	HD\,	54662	A	&O7\,V((f))z\,var?&	&	&	&	\\				
39		&	HD\,	54662	B	&&	&	&	&	\\				
40		&	HD\,	55879		&O9.7\,III&	5.5	&	20	&	10.28	&	0.78	\\
41		&	HD\,	66788		&O8\,V&	3.54	&	21	&	6.52	&	0.44	\\
42		&	HD\,	66811		&O4\,If &	2.12	&	40	&	12.4	&	0.56	\\
43		&	HD\,	69106		&O9.7\,In&	6.36	&	20	&	12.33	&	0.82	\\
44		&	HD\,	93028		&O9\,IV&	&	&	&	\\				
45	$^\dagger$	&	HD\,	93250		&O4\,IIIfc:&	1.62	&	45	&	12.02	&	0.47	\\
46		&	HD\,	149038		&O9.7\,Iab&	5.16	&	24	&	16.45	&	0.86	\\
47		&	HD\,	149757		& O9.5\,Ivnn&	5.84	&	20	&	9.7	&	0.71	\\
48		&	HD\,	151804		&O8\,Iaf&	2.1	&	40	&	28.42	&	0.81	\\
49		&	HD\,	152233		& O6\,III(f)&	&	&	&	\\				
50		&	HD\,	152247		&O9.5\,II-III&	4.88	&	22	&	10.25	&	0.77	\\
51		&	HD\,	152249		&OC9\,Iab&	3.62	&	31	&	18.65	&	0.81	\\
52		&	HD\,	152408		&O8\,Iafpe&	&	&	&	\\				
53	SB2	&	HD\,	153426		&O9\,II-III&	3.98	&	21.6	&	8.16	&	0.66	\\
54		&	HD\,	153919		&O6\,Iaf&	1.82	&	50	&	20.76	&	0.68	\\
55		&	HD\,	154368		&O9.5\,Iab&	3.62	&	31	&	18.65	&	0.81	\\
56		&	HD\,	154643		&O9.5\,III&	5.46	&	20	&	10.28	&	0.78	\\
57		&	HD\,	155806		&O7.5\,V&	3.02	&	23	&	6.83	&	0.44	\\
58	SB2	&	HD\,	155889		&O9.5\,IV&	3.92	&	19.2	&	6.18	&	0.48	\\
59		&	HD\,	156154		&O7.5\,Ibf&	3.88	&	28	&	13.29	&	0.75	\\
60		&	HD\,	162978		&O8\,II(f)&	3.7	&	30	&	13.41	&	0.72	\\
61		&	HD\,	164492		&O7.5\,Vz&	0.62	&	26	&	6.65	&	0.04	\\
62	SB2	&	HD\,	164794	A	&O3.5\,V((f*))&	&	&	&	\\				
63		&	HD\,	164794	B	&O5-5.5\,V((f))&	&	&	&	\\				
64	SB2	&	HD\,	165052	A	&O5.5Vz&	&	&	&	\\				
65		&	HD\,	165052	B	&O8V&	&	&	&	\\				
66		&	HD\,	167263		&O9.5\,II-IIIn&	5.18	&	22	&	12.07	&	0.79	\\
67		&	HD\,	167264		&O9.7\,Iab&	4.34	&	28	&	21.56	&	0.87	\\
68	$^\dagger$	&	HD\,	167771		&O7\,III(f)&	3.52	&	29	&	11.79	&	0.7	\\
69		&	HD\,	186980		&O7.5\,III((f))&	3.52	&	29	&	11.79	&	0.7	\\
70		&	HD\,	188001		&O7.5\,Iabf &	3.38	&	34	&	16.24	&	0.76	\\
71		&	HD\,	188209		&O9.5\,Iab&	3.92	&	30	&	19.26	&	0.83	\\
72		&	HD\,	189957		&O9.7\,III&	5.48	&	20	&	10.28	&	0.78	\\
73	$^\dagger$	&	HD\,	190918		&WN5o+O9\,I&	&	&	&	\\				
74	SB2	&	HD\,	191201	A	&O9.5\,III&	&	&	&	\\				
75		&	HD\,	191201	B	&B0\,IV&	&	&	&	\\				
76		&	HD\,	192281		& O4.5\,Vn(f)&	2.4	&	37	&	11.8	&	0.59	\\
77		&	HD\,	192639		&O7.5\,Iabf&	2.96	&	38	&	17.92	&	0.74	\\
78	SB2	&	HD\,	193322	Aa	&O9\,IV(n)&	&	&	&	\\				
79		&	HD\,	193322	Ab1	&&	&	&	&	\\				
80	SB2	&	HD\,	193443	A	&O9\,III&	4.88	&	21.8	&	10.25	&	0.77	\\
81		&	HD\,	193443	B	&&	&	&	&	\\				
82	$^\dagger$	&	HD\,	199579		&O6.5\,V((f))z&	1.18	&	35	&	8.02	&	0.13	\\
83		&	HD\,	201345		& ON9.5\,IV&	3.7	&	20	&	6.53	&	0.44	\\
84		&	HD\,	203064		&O7.5\,IIIn((f))&	4.04	&	25	&	10.93	&	0.69	\\
85	SB2	&	HD\,	204827		&O9.7\,III&	&	&	&	\\				
86		&	HD\,	206183		&O9.5\,IV-V&	3.52	&	18	&	5.86	&	0.3	\\
87	$^\dagger$	&	HD\,	206267		&O6.5\,V((f))&	&	&	&	\\				
88		&	HD\,	207198		&O9\,II&	4.36	&	25	&	12.42	&	0.76	\\
89		&	HD\,	207538		&O9.7\,IV&	5.92	&	17	&	7.72	&	0.71	\\
90	SB2	&	HD\,	209481	A	&O9\,IV(n)\,var&	&	&	&	\\				
91		&	HD\,	209481	B	&&	&	&	&	\\				
92		&	HD\,	209975		&O9\,Ib&	4.46	&	25	&	15.19	&	0.82	\\
93		&	HD\,	210809		&O9\,Iab&	4.46	&	25	&	15.19	&	0.82	\\
94		&	HD\,	210839		&O6.5\,I(n)fp&	3.18	&	34	&	13.41	&	0.68	\\
95		&	HD\,	214680		&O9\,V&	3.22	&	21	&	6.45	&	0.36	\\
96		&	HD\,	218195		&O8.5\,III&	3.98	&	22	&	8.16	&	0.66	\\
97		&	HD\,	218915		&O9.5\,Iab&	4.18	&	28	&	18.12	&	0.83	\\
98		&	HD\,	227757		&O9.5\,V&	3.72	&	20	&	6.53	&	0.44	\\
99		&	HD\,	258691		& O9.5\,IV&	3.92	&	19	&	6.18	&	0.48	\\
100		&	HD\,	328856		&O9.7\,II&	5.22	&	23	&	13.16	&	0.82	\\
101		&	BD\,$-$13 4930		&O9.7\,V&	3.52	&	18	&	5.86	&	0.3	\\
102		&	BD\,$+$60 499		&O9.5\,V&	3.72	&	20	&	6.53	&	0.44	\\

\hline
\end{longtable}

}
\twocolumn

\appendix

\section{Line profile models for SB2 systems}
\label{sec:apSB2}

In this appendix, we illustrate the intensity model fit for the SB2 system to provide a reference for the primary and secondary designation adopted. 

For HD 37043, our single LSD profile yields various possible EW ratio for the primary and secondary, all compatible with the results of \citet{1987A&A...184..185S} and \citet{2001ApJ...554..362B}. We opt for a solution with a smaller EW for the primary than presented in \citetalias{2017MNRAS.465.2432G} -- this yields a more conservative estimate of the field upper limits. 

As discussed in \citetalias{2017MNRAS.465.2432G}, our LSD profiles for the SB2 HD 164794 (9 Sgr) do not show two components. However, the line characteristics have been well constrained by \citet{2012A&A...542A..95R}, so we here use their broadening value and radial velocity solution to fit our profiles. 

In \citetalias{2017MNRAS.465.2432G}, HD 191201 is listed as an O+O system, but is given a spectral type of O9.5\,III+
B0\,IV. We include the secondary star in our sample.

%------
\begin{figure}
	\begin{center}
\includegraphics[width=0.45\textwidth]{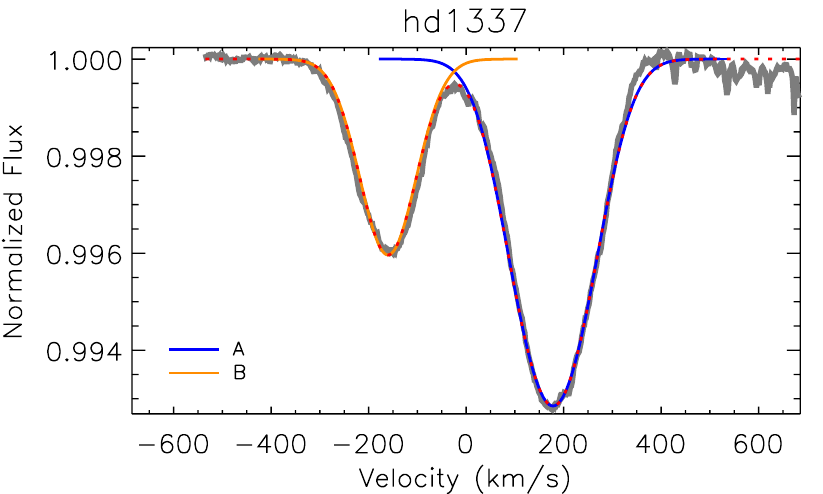}
	\caption{\label{fig:bin} Illustration of model fit for the SB2 systems in the sample. The primary and secondary designation matches those of Table\,\ref{tab:stars}. }
	\end{center}
\end{figure}
%%------

%------
\begin{figure}
	\begin{center}
\includegraphics[width=0.45\textwidth]{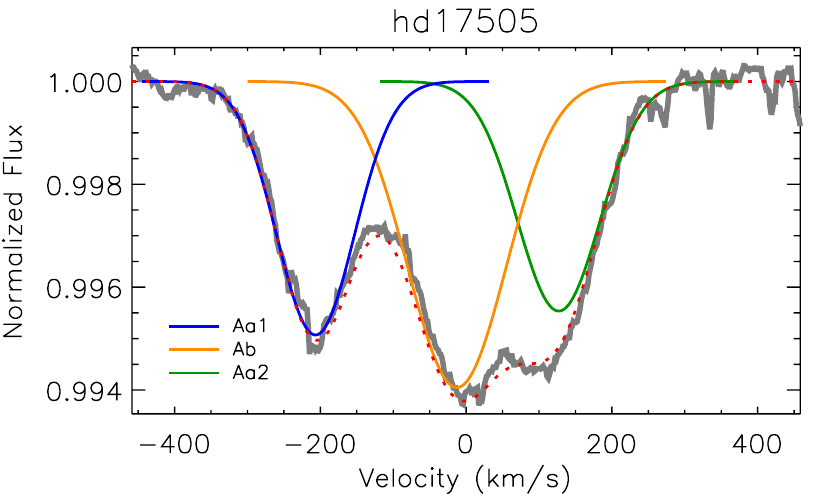}
	\caption{\label{} Same as Fig.\,\ref{fig:bin}  }
	\end{center}
\end{figure}
%%------

%------
\begin{figure}
	\begin{center}
\includegraphics[width=0.45\textwidth]{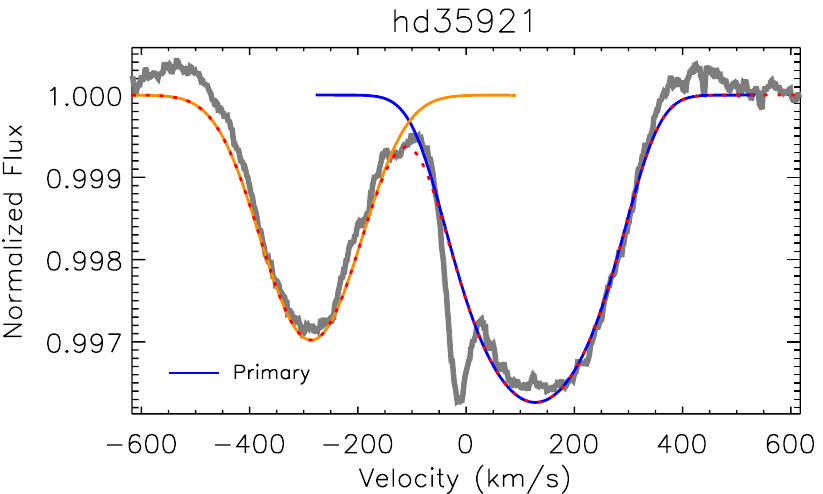}
	\caption{\label{} Same as Fig.\,\ref{fig:bin} }
	\end{center}
\end{figure}
%%------

%------
\begin{figure}
	\begin{center}
\includegraphics[width=0.45\textwidth]{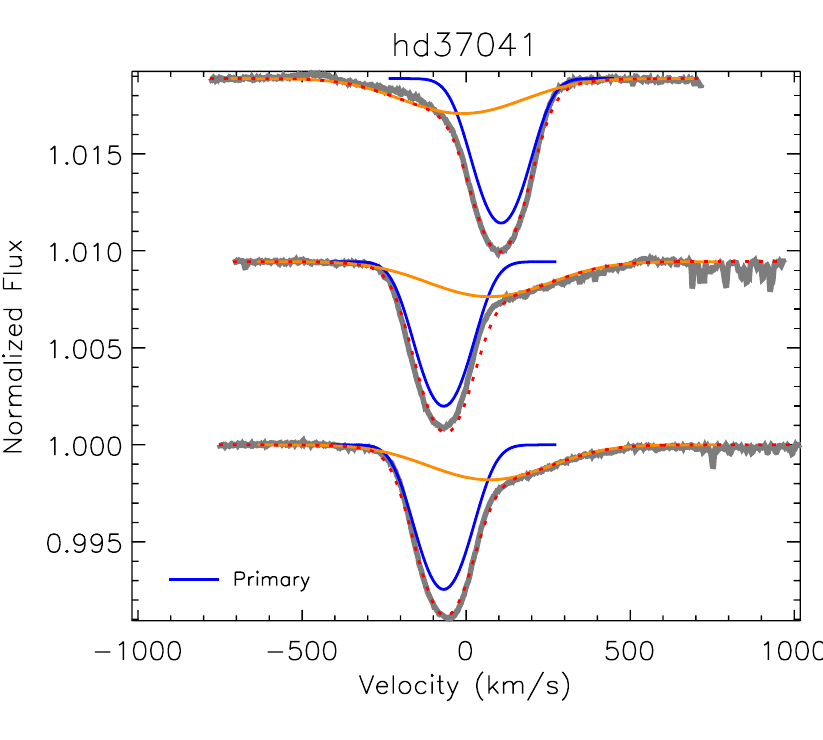}
	\caption{\label{}  Same as Fig.\,\ref{fig:bin}}
	\end{center}
\end{figure}
%%------

%------
\begin{figure}
	\begin{center}
\includegraphics[width=0.45\textwidth]{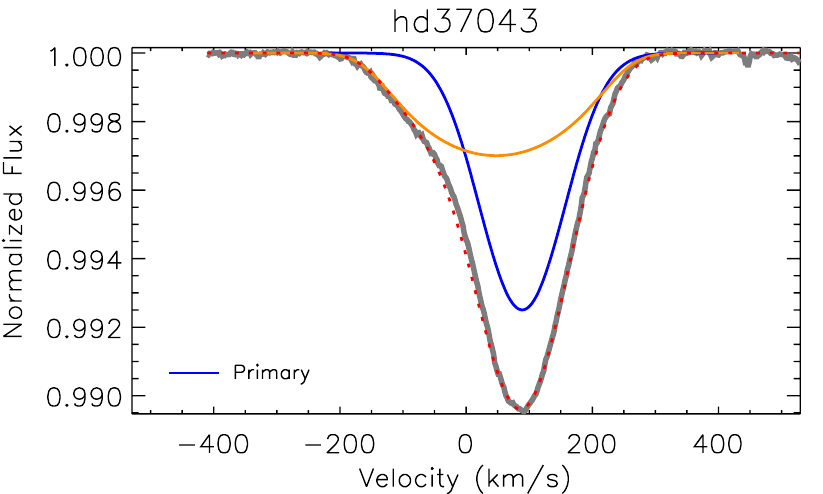}
	\caption{\label{} Same as Fig.\,\ref{fig:bin} }
	\end{center}
\end{figure}
%%------

%------
\begin{figure}
	\begin{center}
\includegraphics[width=0.45\textwidth]{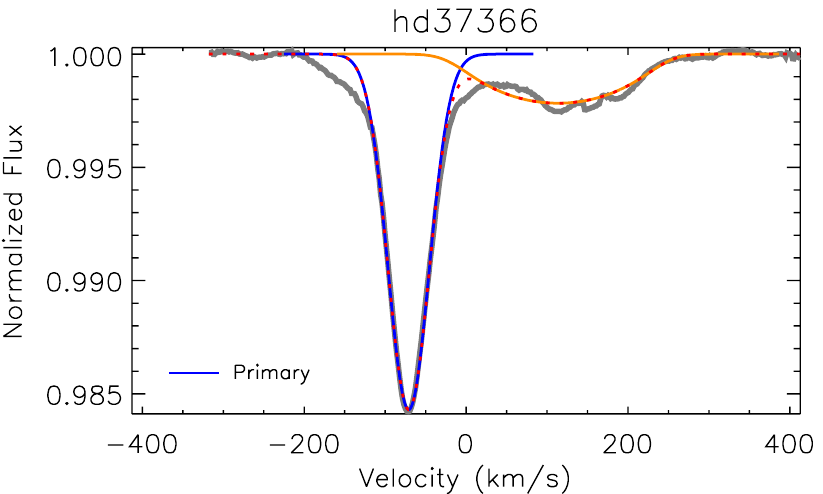}
	\caption{\label{} Same as Fig.\,\ref{fig:bin} }
	\end{center}
\end{figure}
%%------

%------
\begin{figure}
	\begin{center}
\includegraphics[width=0.45\textwidth]{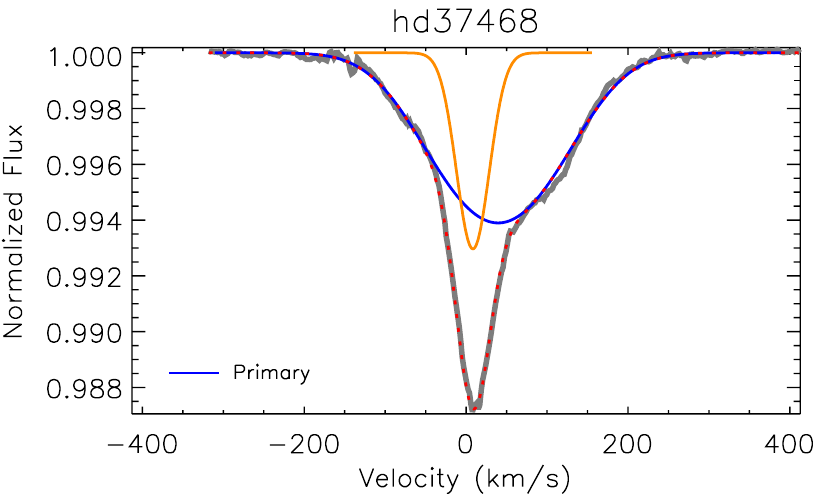}
	\caption{\label{} Same as Fig.\,\ref{fig:bin} }
	\end{center}
\end{figure}
%%------

%------
\begin{figure}
	\begin{center}
\includegraphics[width=0.45\textwidth]{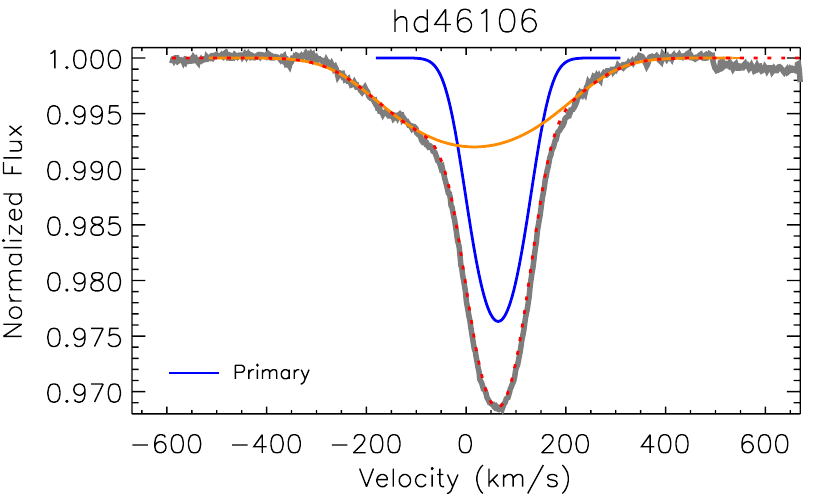}
	\caption{\label{}  Same as Fig.\,\ref{fig:bin}}
	\end{center}
\end{figure}
%%------

%------
\begin{figure}
	\begin{center}
\includegraphics[width=0.45\textwidth]{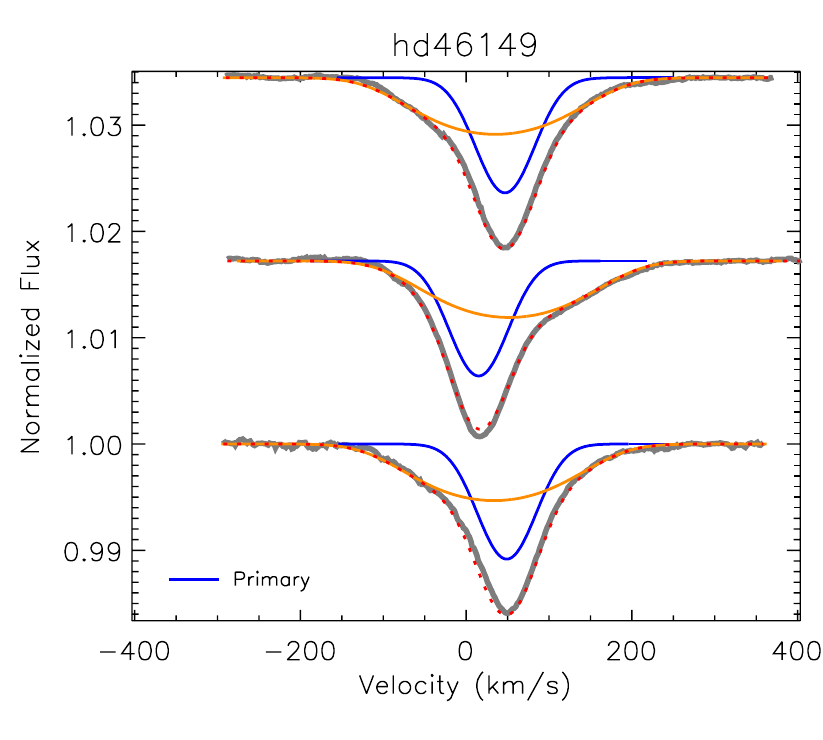}
	\caption{\label{} Same as Fig.\,\ref{fig:bin} }
	\end{center}
\end{figure}
%%------

%------
\begin{figure}
	\begin{center}
\includegraphics[width=0.45\textwidth]{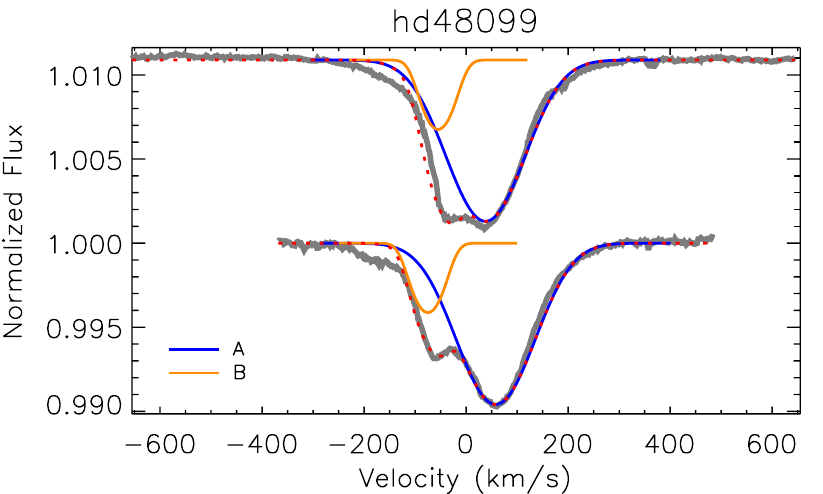}
	\caption{\label{} Same as Fig.\,\ref{fig:bin} }
	\end{center}
\end{figure}
%%------

%------
\begin{figure}
	\begin{center}
\includegraphics[width=0.45\textwidth]{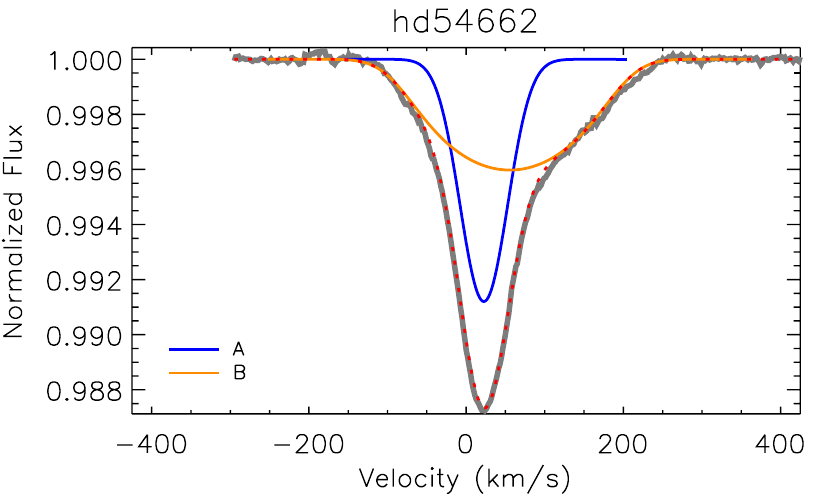}
	\caption{\label{} Same as Fig.\,\ref{fig:bin} }
	\end{center}
\end{figure}
%%------

%------
\begin{figure}
	\begin{center}
\includegraphics[width=0.45\textwidth]{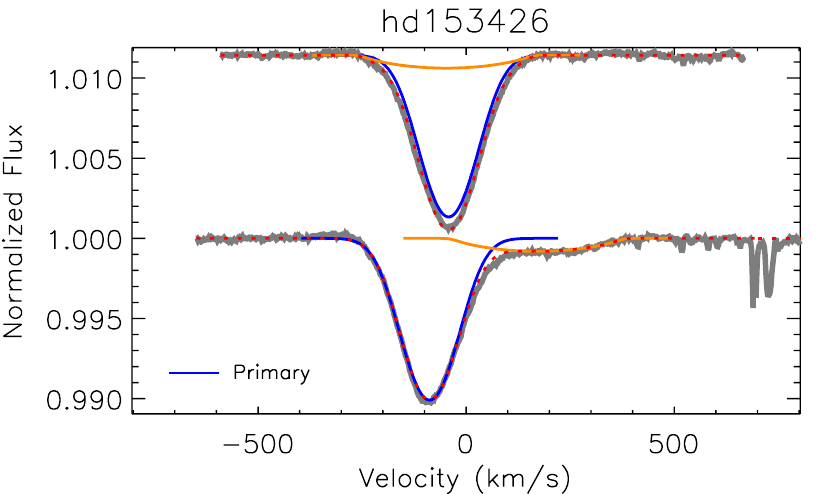}
	\caption{\label{}  Same as Fig.\,\ref{fig:bin}}
	\end{center}
\end{figure}
%%------

%------
\begin{figure}
	\begin{center}
\includegraphics[width=0.45\textwidth]{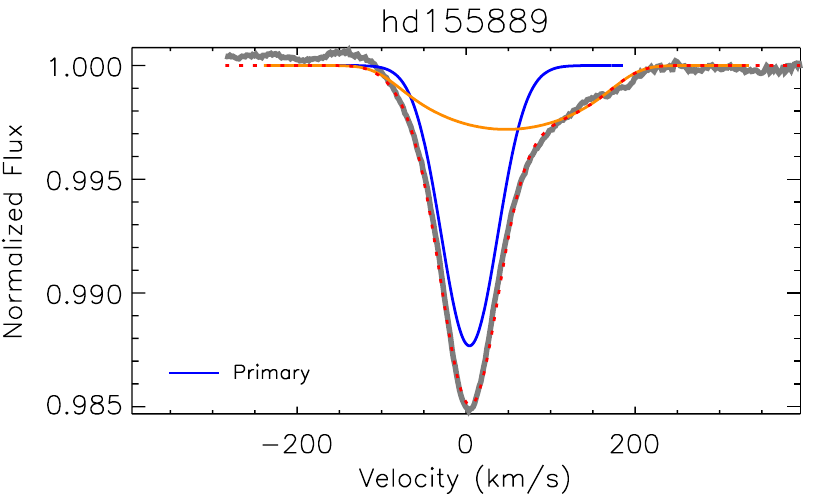}
	\caption{\label{} Same as Fig.\,\ref{fig:bin} }
	\end{center}
\end{figure}
%%------

%------
\begin{figure}
	\begin{center}
\includegraphics[width=0.45\textwidth]{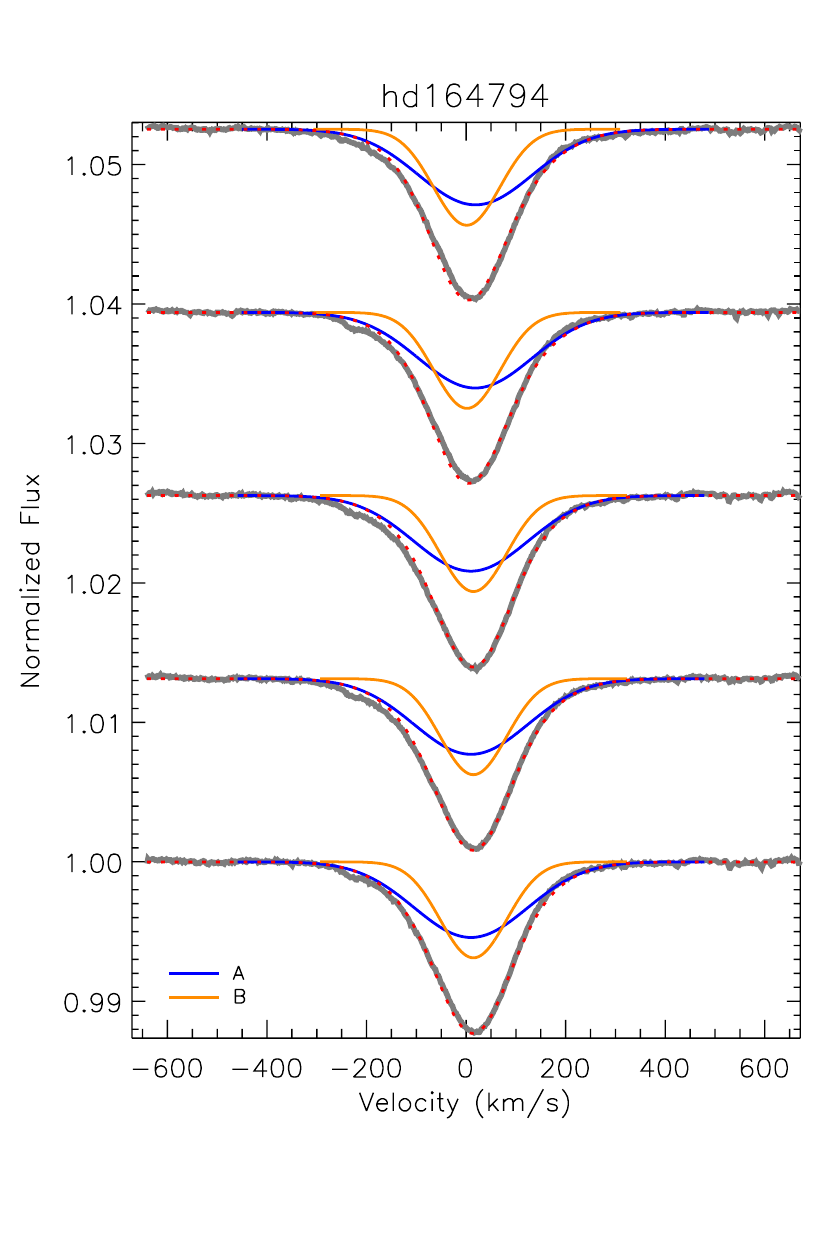}
	\caption{\label{} Same as Fig.\,\ref{fig:bin} }
	\end{center}
\end{figure}
%%------

%------
\begin{figure}
	\begin{center}
\includegraphics[width=0.45\textwidth]{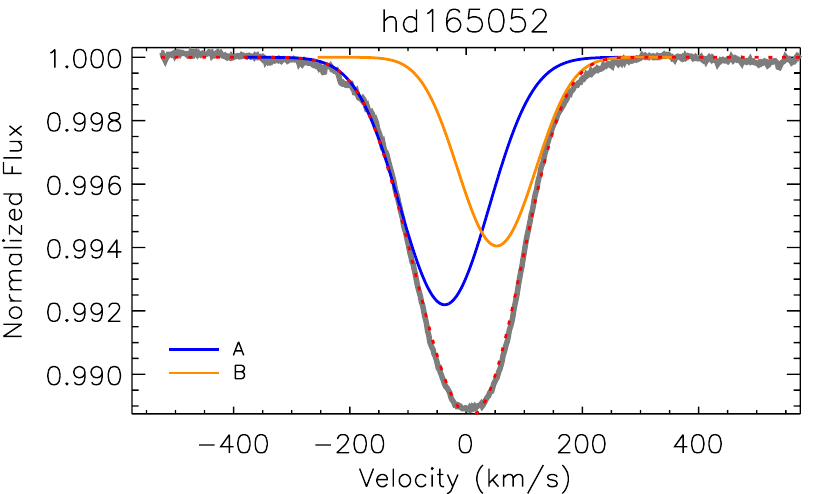}
	\caption{\label{} Same as Fig.\,\ref{fig:bin} }
	\end{center}
\end{figure}
%%------

%------
\begin{figure}
	\begin{center}
\includegraphics[width=0.45\textwidth]{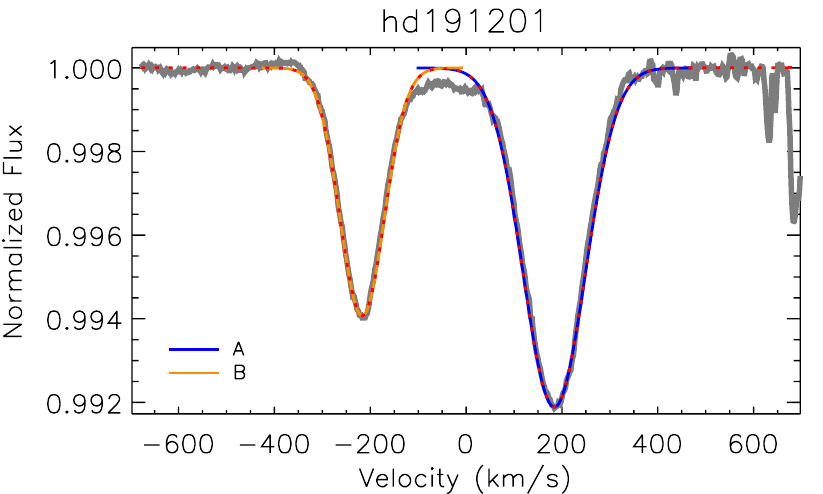}
	\caption{\label{} Same as Fig.\,\ref{fig:bin} }
	\end{center}
\end{figure}
%%------

%------
\begin{figure}
	\begin{center}
\includegraphics[width=0.45\textwidth]{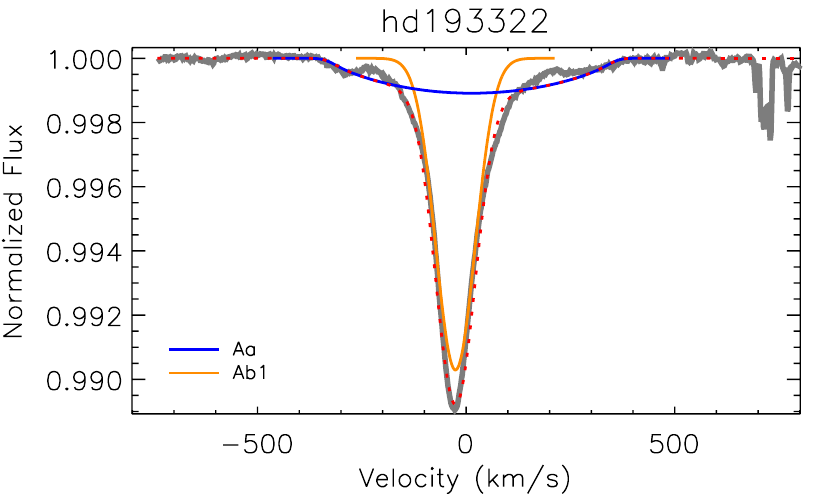}
	\caption{\label{} Same as Fig.\,\ref{fig:bin} }
	\end{center}
\end{figure}
%%------

%------
\begin{figure}
	\begin{center}
\includegraphics[width=0.45\textwidth]{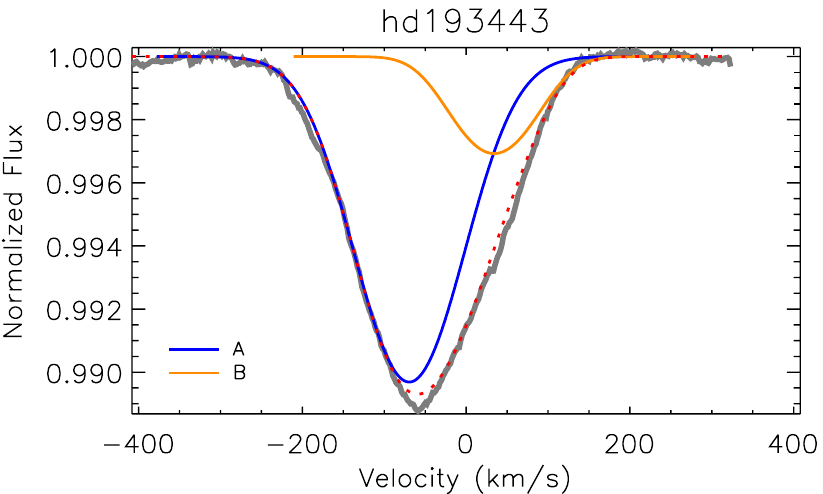}
	\caption{\label{} Same as Fig.\,\ref{fig:bin} }
	\end{center}
\end{figure}
%%------

%------
\begin{figure}
	\begin{center}
\includegraphics[width=0.45\textwidth]{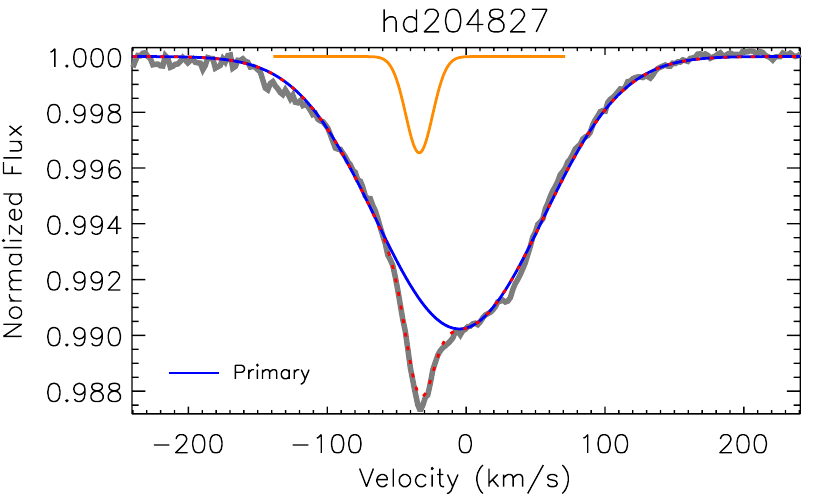}
	\caption{\label{} Same as Fig.\,\ref{fig:bin} }
	\end{center}
\end{figure}
%%------

%------
\begin{figure}
	\begin{center}
\includegraphics[width=0.45\textwidth]{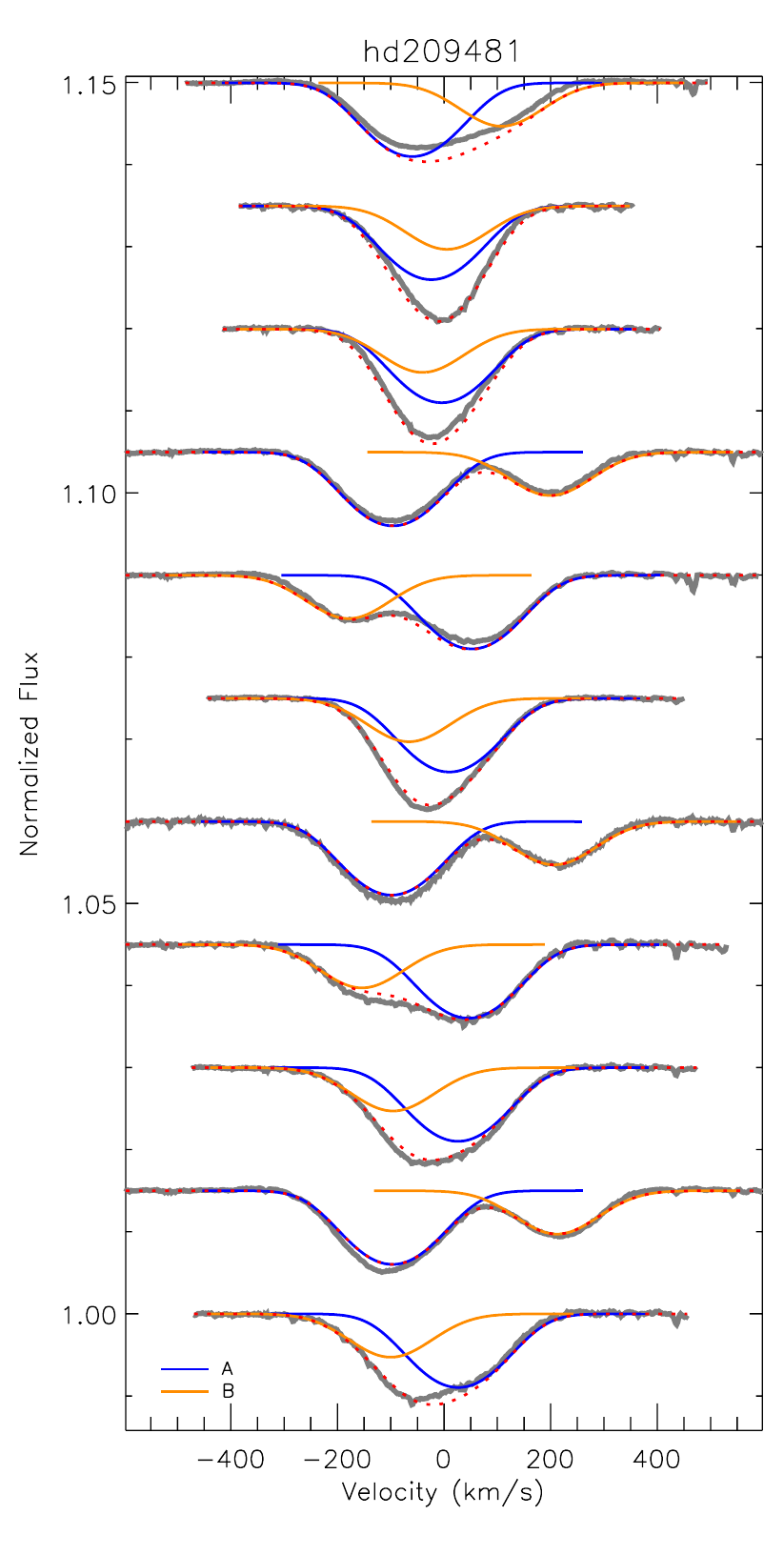}
	\caption{\label{} Same as Fig.\,\ref{fig:bin} }
	\end{center}
\end{figure}
%%------

\label{lastpage}
\end{document}